\documentclass[12pt,a4paper]{article}
\usepackage[a4paper, hmargin=2cm, vmargin=2cm]{geometry}
\usepackage{authblk}
\usepackage[dvipsnames]{xcolor}
\usepackage[pdftex,colorlinks]{hyperref}
\usepackage[all]{hypcap}
\usepackage{amssymb,amsmath}
\usepackage{mathcomp}
\usepackage{color} 
\usepackage{stmaryrd}
\usepackage{IEEEtrantools}
\usepackage{cite}
\usepackage{graphicx}
\usepackage{setspace}
\usepackage[colorinlistoftodos,prependcaption,textsize=footnotesize]{todonotes}		
\usepackage{verbatim}		
\usepackage{courier}
\usepackage{mdframed}
\usepackage{tabularx,longtable,multirow,subfigure,caption} 
\usepackage{commath} 
\usepackage{epstopdf}
\usepackage{float}
\usepackage{bm}
\usepackage{textpos}
\usepackage{url} 
\usepackage[english]{babel}
\usepackage{dsfont}
\usepackage{array}
\usepackage{latexsym}
\usepackage{mathtools}
\usepackage{afterpage}
\usepackage{algorithm}
\usepackage{algorithmic}
\usepackage[capitalise]{cleveref} 
\usepackage{soul}
\usepackage{tikz}
\usepackage{xargs}
\usepackage{keyval}
\usepackage{pgfkeys}
\usepackage{fix-cm}
\usepackage{tcolorbox}
\usepackage{colortbl}

\definecolor{AliceBlue}{rgb}{0.94,0.97,1.00}
\definecolor{AntiqueWhite1}{rgb}{1.00,0.94,0.86}
\definecolor{AntiqueWhite2}{rgb}{0.93,0.87,0.80}
\definecolor{AntiqueWhite3}{rgb}{0.80,0.75,0.69}
\definecolor{AntiqueWhite4}{rgb}{0.55,0.51,0.47}
\definecolor{AntiqueWhite}{rgb}{0.98,0.92,0.84}
\definecolor{BlanchedAlmond}{rgb}{1.00,0.92,0.80}
\definecolor{BlueViolet}{rgb}{0.54,0.17,0.89}
\definecolor{CadetBlue1}{rgb}{0.60,0.96,1.00}
\definecolor{CadetBlue2}{rgb}{0.56,0.90,0.93}
\definecolor{CadetBlue3}{rgb}{0.48,0.77,0.80}
\definecolor{CadetBlue4}{rgb}{0.33,0.53,0.55}
\definecolor{CadetBlue}{rgb}{0.37,0.62,0.63}
\definecolor{CornflowerBlue}{rgb}{0.39,0.58,0.93}
\definecolor{DarkBlue}{rgb}{0.00,0.00,0.55}
\definecolor{DarkCyan}{rgb}{0.00,0.55,0.55}
\definecolor{DarkGoldenrod1}{rgb}{1.00,0.73,0.06}
\definecolor{DarkGoldenrod2}{rgb}{0.93,0.68,0.05}
\definecolor{DarkGoldenrod3}{rgb}{0.80,0.58,0.05}
\definecolor{DarkGoldenrod4}{rgb}{0.55,0.40,0.03}
\definecolor{DarkGoldenrod}{rgb}{0.72,0.53,0.04}
\definecolor{DarkGray}{rgb}{0.66,0.66,0.66}
\definecolor{DarkGreen}{rgb}{0.00,0.39,0.00}
\definecolor{DarkGrey}{rgb}{0.66,0.66,0.66}
\definecolor{DarkKhaki}{rgb}{0.74,0.72,0.42}
\definecolor{DarkMagenta}{rgb}{0.55,0.00,0.55}
\definecolor{DarkOliveGreen1}{rgb}{0.79,1.00,0.44}
\definecolor{DarkOliveGreen2}{rgb}{0.74,0.93,0.41}
\definecolor{DarkOliveGreen3}{rgb}{0.64,0.80,0.35}
\definecolor{DarkOliveGreen4}{rgb}{0.43,0.55,0.24}
\definecolor{DarkOliveGreen}{rgb}{0.33,0.42,0.18}
\definecolor{DarkOrange1}{rgb}{1.00,0.50,0.00}
\definecolor{DarkOrange2}{rgb}{0.93,0.46,0.00}
\definecolor{DarkOrange3}{rgb}{0.80,0.40,0.00}
\definecolor{DarkOrange4}{rgb}{0.55,0.27,0.00}
\definecolor{DarkOrange}{rgb}{1.00,0.55,0.00}
\definecolor{DarkOrchid1}{rgb}{0.75,0.24,1.00}
\definecolor{DarkOrchid2}{rgb}{0.70,0.23,0.93}
\definecolor{DarkOrchid3}{rgb}{0.60,0.20,0.80}
\definecolor{DarkOrchid4}{rgb}{0.41,0.13,0.55}
\definecolor{DarkOrchid}{rgb}{0.60,0.20,0.80}
\definecolor{DarkRed}{rgb}{0.55,0.00,0.00}
\definecolor{DarkSalmon}{rgb}{0.91,0.59,0.48}
\definecolor{DarkSeaGreen1}{rgb}{0.76,1.00,0.76}
\definecolor{DarkSeaGreen2}{rgb}{0.71,0.93,0.71}
\definecolor{DarkSeaGreen3}{rgb}{0.61,0.80,0.61}
\definecolor{DarkSeaGreen4}{rgb}{0.41,0.55,0.41}
\definecolor{DarkSeaGreen}{rgb}{0.56,0.74,0.56}
\definecolor{DarkSlateBlue}{rgb}{0.28,0.24,0.55}
\definecolor{DarkSlateGray1}{rgb}{0.59,1.00,1.00}
\definecolor{DarkSlateGray2}{rgb}{0.55,0.93,0.93}
\definecolor{DarkSlateGray3}{rgb}{0.47,0.80,0.80}
\definecolor{DarkSlateGray4}{rgb}{0.32,0.55,0.55}
\definecolor{DarkSlateGray}{rgb}{0.18,0.31,0.31}
\definecolor{DarkSlateGrey}{rgb}{0.18,0.31,0.31}
\definecolor{DarkTurquoise}{rgb}{0.00,0.81,0.82}
\definecolor{DarkViolet}{rgb}{0.58,0.00,0.83}
\definecolor{DeepPink1}{rgb}{1.00,0.08,0.58}
\definecolor{DeepPink2}{rgb}{0.93,0.07,0.54}
\definecolor{DeepPink3}{rgb}{0.80,0.06,0.46}
\definecolor{DeepPink4}{rgb}{0.55,0.04,0.31}
\definecolor{DeepPink}{rgb}{1.00,0.08,0.58}
\definecolor{DeepSkyBlue1}{rgb}{0.00,0.75,1.00}
\definecolor{DeepSkyBlue2}{rgb}{0.00,0.70,0.93}
\definecolor{DeepSkyBlue3}{rgb}{0.00,0.60,0.80}
\definecolor{DeepSkyBlue4}{rgb}{0.00,0.41,0.55}
\definecolor{DeepSkyBlue}{rgb}{0.00,0.75,1.00}
\definecolor{DimGray}{rgb}{0.41,0.41,0.41}
\definecolor{DimGrey}{rgb}{0.41,0.41,0.41}
\definecolor{DodgerBlue1}{rgb}{0.12,0.56,1.00}
\definecolor{DodgerBlue2}{rgb}{0.11,0.53,0.93}
\definecolor{DodgerBlue3}{rgb}{0.09,0.45,0.80}
\definecolor{DodgerBlue4}{rgb}{0.06,0.31,0.55}
\definecolor{DodgerBlue}{rgb}{0.12,0.56,1.00}
\definecolor{FloralWhite}{rgb}{1.00,0.98,0.94}
\definecolor{ForestGreen}{rgb}{0.13,0.55,0.13}
\definecolor{GhostWhite}{rgb}{0.97,0.97,1.00}
\definecolor{GreenYellow}{rgb}{0.68,1.00,0.18}
\definecolor{HotPink1}{rgb}{1.00,0.43,0.71}
\definecolor{HotPink2}{rgb}{0.93,0.42,0.65}
\definecolor{HotPink3}{rgb}{0.80,0.38,0.56}
\definecolor{HotPink4}{rgb}{0.55,0.23,0.38}
\definecolor{HotPink}{rgb}{1.00,0.41,0.71}
\definecolor{IndianRed1}{rgb}{1.00,0.42,0.42}
\definecolor{IndianRed2}{rgb}{0.93,0.39,0.39}
\definecolor{IndianRed3}{rgb}{0.80,0.33,0.33}
\definecolor{IndianRed4}{rgb}{0.55,0.23,0.23}
\definecolor{IndianRed}{rgb}{0.80,0.36,0.36}
\definecolor{LavenderBlush1}{rgb}{1.00,0.94,0.96}
\definecolor{LavenderBlush2}{rgb}{0.93,0.88,0.90}
\definecolor{LavenderBlush3}{rgb}{0.80,0.76,0.77}
\definecolor{LavenderBlush4}{rgb}{0.55,0.51,0.53}
\definecolor{LavenderBlush}{rgb}{1.00,0.94,0.96}
\definecolor{LawnGreen}{rgb}{0.49,0.99,0.00}
\definecolor{LemonChiffon1}{rgb}{1.00,0.98,0.80}
\definecolor{LemonChiffon2}{rgb}{0.93,0.91,0.75}
\definecolor{LemonChiffon3}{rgb}{0.80,0.79,0.65}
\definecolor{LemonChiffon4}{rgb}{0.55,0.54,0.44}
\definecolor{LemonChiffon}{rgb}{1.00,0.98,0.80}
\definecolor{LightBlue1}{rgb}{0.75,0.94,1.00}
\definecolor{LightBlue2}{rgb}{0.70,0.87,0.93}
\definecolor{LightBlue3}{rgb}{0.60,0.75,0.80}
\definecolor{LightBlue4}{rgb}{0.41,0.51,0.55}
\definecolor{LightBlue}{rgb}{0.68,0.85,0.90}
\definecolor{LightCoral}{rgb}{0.94,0.50,0.50}
\definecolor{LightCyan1}{rgb}{0.88,1.00,1.00}
\definecolor{LightCyan2}{rgb}{0.82,0.93,0.93}
\definecolor{LightCyan3}{rgb}{0.71,0.80,0.80}
\definecolor{LightCyan4}{rgb}{0.48,0.55,0.55}
\definecolor{LightCyan}{rgb}{0.88,1.00,1.00}
\definecolor{LightGoldenrod1}{rgb}{1.00,0.93,0.55}
\definecolor{LightGoldenrod2}{rgb}{0.93,0.86,0.51}
\definecolor{LightGoldenrod3}{rgb}{0.80,0.75,0.44}
\definecolor{LightGoldenrod4}{rgb}{0.55,0.51,0.30}
\definecolor{LightGoldenrodYellow}{rgb}{0.98,0.98,0.82}
\definecolor{LightGoldenrod}{rgb}{0.93,0.87,0.51}
\definecolor{LightGray}{rgb}{0.83,0.83,0.83}
\definecolor{LightGreen}{rgb}{0.56,0.93,0.56}
\definecolor{LightGrey}{rgb}{0.83,0.83,0.83}
\definecolor{LightPink1}{rgb}{1.00,0.68,0.73}
\definecolor{LightPink2}{rgb}{0.93,0.64,0.68}
\definecolor{LightPink3}{rgb}{0.80,0.55,0.58}
\definecolor{LightPink4}{rgb}{0.55,0.37,0.40}
\definecolor{LightPink}{rgb}{1.00,0.71,0.76}
\definecolor{LightSalmon1}{rgb}{1.00,0.63,0.48}
\definecolor{LightSalmon2}{rgb}{0.93,0.58,0.45}
\definecolor{LightSalmon3}{rgb}{0.80,0.51,0.38}
\definecolor{LightSalmon4}{rgb}{0.55,0.34,0.26}
\definecolor{LightSalmon}{rgb}{1.00,0.63,0.48}
\definecolor{LightSeaGreen}{rgb}{0.13,0.70,0.67}
\definecolor{LightSkyBlue1}{rgb}{0.69,0.89,1.00}
\definecolor{LightSkyBlue2}{rgb}{0.64,0.83,0.93}
\definecolor{LightSkyBlue3}{rgb}{0.55,0.71,0.80}
\definecolor{LightSkyBlue4}{rgb}{0.38,0.48,0.55}
\definecolor{LightSkyBlue}{rgb}{0.53,0.81,0.98}
\definecolor{LightSlateBlue}{rgb}{0.52,0.44,1.00}
\definecolor{LightSlateGray}{rgb}{0.47,0.53,0.60}
\definecolor{LightSlateGrey}{rgb}{0.47,0.53,0.60}
\definecolor{LightSteelBlue1}{rgb}{0.79,0.88,1.00}
\definecolor{LightSteelBlue2}{rgb}{0.74,0.82,0.93}
\definecolor{LightSteelBlue3}{rgb}{0.64,0.71,0.80}
\definecolor{LightSteelBlue4}{rgb}{0.43,0.48,0.55}
\definecolor{LightSteelBlue}{rgb}{0.69,0.77,0.87}
\definecolor{LightYellow1}{rgb}{1.00,1.00,0.88}
\definecolor{LightYellow2}{rgb}{0.93,0.93,0.82}
\definecolor{LightYellow3}{rgb}{0.80,0.80,0.71}
\definecolor{LightYellow4}{rgb}{0.55,0.55,0.48}
\definecolor{LightYellow}{rgb}{1.00,1.00,0.88}
\definecolor{LimeGreen}{rgb}{0.20,0.80,0.20}
\definecolor{MediumAquamarine}{rgb}{0.40,0.80,0.67}
\definecolor{MediumBlue}{rgb}{0.00,0.00,0.80}
\definecolor{MediumOrchid1}{rgb}{0.88,0.40,1.00}
\definecolor{MediumOrchid2}{rgb}{0.82,0.37,0.93}
\definecolor{MediumOrchid3}{rgb}{0.71,0.32,0.80}
\definecolor{MediumOrchid4}{rgb}{0.48,0.22,0.55}
\definecolor{MediumOrchid}{rgb}{0.73,0.33,0.83}
\definecolor{MediumPurple1}{rgb}{0.67,0.51,1.00}
\definecolor{MediumPurple2}{rgb}{0.62,0.47,0.93}
\definecolor{MediumPurple3}{rgb}{0.54,0.41,0.80}
\definecolor{MediumPurple4}{rgb}{0.36,0.28,0.55}
\definecolor{MediumPurple}{rgb}{0.58,0.44,0.86}
\definecolor{MediumSeaGreen}{rgb}{0.24,0.70,0.44}
\definecolor{MediumSlateBlue}{rgb}{0.48,0.41,0.93}
\definecolor{MediumSpringGreen}{rgb}{0.00,0.98,0.60}
\definecolor{MediumTurquoise}{rgb}{0.28,0.82,0.80}
\definecolor{MediumVioletRed}{rgb}{0.78,0.08,0.52}
\definecolor{MidnightBlue}{rgb}{0.10,0.10,0.44}
\definecolor{MintCream}{rgb}{0.96,1.00,0.98}
\definecolor{MistyRose1}{rgb}{1.00,0.89,0.88}
\definecolor{MistyRose2}{rgb}{0.93,0.84,0.82}
\definecolor{MistyRose3}{rgb}{0.80,0.72,0.71}
\definecolor{MistyRose4}{rgb}{0.55,0.49,0.48}
\definecolor{MistyRose}{rgb}{1.00,0.89,0.88}
\definecolor{NavajoWhite1}{rgb}{1.00,0.87,0.68}
\definecolor{NavajoWhite2}{rgb}{0.93,0.81,0.63}
\definecolor{NavajoWhite3}{rgb}{0.80,0.70,0.55}
\definecolor{NavajoWhite4}{rgb}{0.55,0.47,0.37}
\definecolor{NavajoWhite}{rgb}{1.00,0.87,0.68}
\definecolor{NavyBlue}{rgb}{0.00,0.00,0.50}
\definecolor{OldLace}{rgb}{0.99,0.96,0.90}
\definecolor{OliveDrab1}{rgb}{0.75,1.00,0.24}
\definecolor{OliveDrab2}{rgb}{0.70,0.93,0.23}
\definecolor{OliveDrab3}{rgb}{0.60,0.80,0.20}
\definecolor{OliveDrab4}{rgb}{0.41,0.55,0.13}
\definecolor{OliveDrab}{rgb}{0.42,0.56,0.14}
\definecolor{OrangeRed1}{rgb}{1.00,0.27,0.00}
\definecolor{OrangeRed2}{rgb}{0.93,0.25,0.00}
\definecolor{OrangeRed3}{rgb}{0.80,0.22,0.00}
\definecolor{OrangeRed4}{rgb}{0.55,0.15,0.00}
\definecolor{OrangeRed}{rgb}{1.00,0.27,0.00}
\definecolor{PaleGoldenrod}{rgb}{0.93,0.91,0.67}
\definecolor{PaleGreen1}{rgb}{0.60,1.00,0.60}
\definecolor{PaleGreen2}{rgb}{0.56,0.93,0.56}
\definecolor{PaleGreen3}{rgb}{0.49,0.80,0.49}
\definecolor{PaleGreen4}{rgb}{0.33,0.55,0.33}
\definecolor{PaleGreen}{rgb}{0.60,0.98,0.60}
\definecolor{PaleTurquoise1}{rgb}{0.73,1.00,1.00}
\definecolor{PaleTurquoise2}{rgb}{0.68,0.93,0.93}
\definecolor{PaleTurquoise3}{rgb}{0.59,0.80,0.80}
\definecolor{PaleTurquoise4}{rgb}{0.40,0.55,0.55}
\definecolor{PaleTurquoise}{rgb}{0.69,0.93,0.93}
\definecolor{PaleVioletRed1}{rgb}{1.00,0.51,0.67}
\definecolor{PaleVioletRed2}{rgb}{0.93,0.47,0.62}
\definecolor{PaleVioletRed3}{rgb}{0.80,0.41,0.54}
\definecolor{PaleVioletRed4}{rgb}{0.55,0.28,0.36}
\definecolor{PaleVioletRed}{rgb}{0.86,0.44,0.58}
\definecolor{PapayaWhip}{rgb}{1.00,0.94,0.84}
\definecolor{PeachPuff1}{rgb}{1.00,0.85,0.73}
\definecolor{PeachPuff2}{rgb}{0.93,0.80,0.68}
\definecolor{PeachPuff3}{rgb}{0.80,0.69,0.58}
\definecolor{PeachPuff4}{rgb}{0.55,0.47,0.40}
\definecolor{PeachPuff}{rgb}{1.00,0.85,0.73}
\definecolor{PowderBlue}{rgb}{0.69,0.88,0.90}
\definecolor{RosyBrown1}{rgb}{1.00,0.76,0.76}
\definecolor{RosyBrown2}{rgb}{0.93,0.71,0.71}
\definecolor{RosyBrown3}{rgb}{0.80,0.61,0.61}
\definecolor{RosyBrown4}{rgb}{0.55,0.41,0.41}
\definecolor{RosyBrown}{rgb}{0.74,0.56,0.56}
\definecolor{RoyalBlue1}{rgb}{0.28,0.46,1.00}
\definecolor{RoyalBlue2}{rgb}{0.26,0.43,0.93}
\definecolor{RoyalBlue3}{rgb}{0.23,0.37,0.80}
\definecolor{RoyalBlue4}{rgb}{0.15,0.25,0.55}
\definecolor{RoyalBlue}{rgb}{0.25,0.41,0.88}
\definecolor{SaddleBrown}{rgb}{0.55,0.27,0.07}
\definecolor{SandyBrown}{rgb}{0.96,0.64,0.38}
\definecolor{SeaGreen1}{rgb}{0.33,1.00,0.62}
\definecolor{SeaGreen2}{rgb}{0.31,0.93,0.58}
\definecolor{SeaGreen3}{rgb}{0.26,0.80,0.50}
\definecolor{SeaGreen4}{rgb}{0.18,0.55,0.34}
\definecolor{SeaGreen}{rgb}{0.18,0.55,0.34}
\definecolor{SkyBlue1}{rgb}{0.53,0.81,1.00}
\definecolor{SkyBlue2}{rgb}{0.49,0.75,0.93}
\definecolor{SkyBlue3}{rgb}{0.42,0.65,0.80}
\definecolor{SkyBlue4}{rgb}{0.29,0.44,0.55}
\definecolor{SkyBlue}{rgb}{0.53,0.81,0.92}
\definecolor{SlateBlue1}{rgb}{0.51,0.44,1.00}
\definecolor{SlateBlue2}{rgb}{0.48,0.40,0.93}
\definecolor{SlateBlue3}{rgb}{0.41,0.35,0.80}
\definecolor{SlateBlue4}{rgb}{0.28,0.24,0.55}
\definecolor{SlateBlue}{rgb}{0.42,0.35,0.80}
\definecolor{SlateGray1}{rgb}{0.78,0.89,1.00}
\definecolor{SlateGray2}{rgb}{0.73,0.83,0.93}
\definecolor{SlateGray3}{rgb}{0.62,0.71,0.80}
\definecolor{SlateGray4}{rgb}{0.42,0.48,0.55}
\definecolor{SlateGray}{rgb}{0.44,0.50,0.56}
\definecolor{SlateGrey}{rgb}{0.44,0.50,0.56}
\definecolor{SpringGreen1}{rgb}{0.00,1.00,0.50}
\definecolor{SpringGreen2}{rgb}{0.00,0.93,0.46}
\definecolor{SpringGreen3}{rgb}{0.00,0.80,0.40}
\definecolor{SpringGreen4}{rgb}{0.00,0.55,0.27}
\definecolor{SpringGreen}{rgb}{0.00,1.00,0.50}
\definecolor{SteelBlue1}{rgb}{0.39,0.72,1.00}
\definecolor{SteelBlue2}{rgb}{0.36,0.67,0.93}
\definecolor{SteelBlue3}{rgb}{0.31,0.58,0.80}
\definecolor{SteelBlue4}{rgb}{0.21,0.39,0.55}
\definecolor{SteelBlue}{rgb}{0.27,0.51,0.71}
\definecolor{VioletRed1}{rgb}{1.00,0.24,0.59}
\definecolor{VioletRed2}{rgb}{0.93,0.23,0.55}
\definecolor{VioletRed3}{rgb}{0.80,0.20,0.47}
\definecolor{VioletRed4}{rgb}{0.55,0.13,0.32}
\definecolor{VioletRed}{rgb}{0.82,0.13,0.56}
\definecolor{WhiteSmoke}{rgb}{0.96,0.96,0.96}
\definecolor{YellowGreen}{rgb}{0.60,0.80,0.20}
\definecolor{aliceblue}{rgb}{0.94,0.97,1.00}
\definecolor{antiquewhite}{rgb}{0.98,0.92,0.84}
\definecolor{aquamarine1}{rgb}{0.50,1.00,0.83}
\definecolor{aquamarine2}{rgb}{0.46,0.93,0.78}
\definecolor{aquamarine3}{rgb}{0.40,0.80,0.67}
\definecolor{aquamarine4}{rgb}{0.27,0.55,0.45}
\definecolor{aquamarine}{rgb}{0.50,1.00,0.83}
\definecolor{azure1}{rgb}{0.94,1.00,1.00}
\definecolor{azure2}{rgb}{0.88,0.93,0.93}
\definecolor{azure3}{rgb}{0.76,0.80,0.80}
\definecolor{azure4}{rgb}{0.51,0.55,0.55}
\definecolor{azure}{rgb}{0.94,1.00,1.00}
\definecolor{beige}{rgb}{0.96,0.96,0.86}
\definecolor{bisque1}{rgb}{1.00,0.89,0.77}
\definecolor{bisque2}{rgb}{0.93,0.84,0.72}
\definecolor{bisque3}{rgb}{0.80,0.72,0.62}
\definecolor{bisque4}{rgb}{0.55,0.49,0.42}
\definecolor{bisque}{rgb}{1.00,0.89,0.77}
\definecolor{black}{rgb}{0.00,0.00,0.00}
\definecolor{blanchedalmond}{rgb}{1.00,0.92,0.80}
\definecolor{blue1}{rgb}{0.00,0.00,1.00}
\definecolor{blue2}{rgb}{0.00,0.00,0.93}
\definecolor{blue3}{rgb}{0.00,0.00,0.80}
\definecolor{blue4}{rgb}{0.00,0.00,0.55}
\definecolor{blueviolet}{rgb}{0.54,0.17,0.89}
\definecolor{blue}{rgb}{0.00,0.00,1.00}
\definecolor{brown1}{rgb}{1.00,0.25,0.25}
\definecolor{brown2}{rgb}{0.93,0.23,0.23}
\definecolor{brown3}{rgb}{0.80,0.20,0.20}
\definecolor{brown4}{rgb}{0.55,0.14,0.14}
\definecolor{brown}{rgb}{0.65,0.16,0.16}
\definecolor{burlywood1}{rgb}{1.00,0.83,0.61}
\definecolor{burlywood2}{rgb}{0.93,0.77,0.57}
\definecolor{burlywood3}{rgb}{0.80,0.67,0.49}
\definecolor{burlywood4}{rgb}{0.55,0.45,0.33}
\definecolor{burlywood}{rgb}{0.87,0.72,0.53}
\definecolor{cadetblue}{rgb}{0.37,0.62,0.63}
\definecolor{chartreuse1}{rgb}{0.50,1.00,0.00}
\definecolor{chartreuse2}{rgb}{0.46,0.93,0.00}
\definecolor{chartreuse3}{rgb}{0.40,0.80,0.00}
\definecolor{chartreuse4}{rgb}{0.27,0.55,0.00}
\definecolor{chartreuse}{rgb}{0.50,1.00,0.00}
\definecolor{chocolate1}{rgb}{1.00,0.50,0.14}
\definecolor{chocolate2}{rgb}{0.93,0.46,0.13}
\definecolor{chocolate3}{rgb}{0.80,0.40,0.11}
\definecolor{chocolate4}{rgb}{0.55,0.27,0.07}
\definecolor{chocolate}{rgb}{0.82,0.41,0.12}
\definecolor{coral1}{rgb}{1.00,0.45,0.34}
\definecolor{coral2}{rgb}{0.93,0.42,0.31}
\definecolor{coral3}{rgb}{0.80,0.36,0.27}
\definecolor{coral4}{rgb}{0.55,0.24,0.18}
\definecolor{coral}{rgb}{1.00,0.50,0.31}
\definecolor{cornflowerblue}{rgb}{0.39,0.58,0.93}
\definecolor{cornsilk1}{rgb}{1.00,0.97,0.86}
\definecolor{cornsilk2}{rgb}{0.93,0.91,0.80}
\definecolor{cornsilk3}{rgb}{0.80,0.78,0.69}
\definecolor{cornsilk4}{rgb}{0.55,0.53,0.47}
\definecolor{cornsilk}{rgb}{1.00,0.97,0.86}
\definecolor{cyan1}{rgb}{0.00,1.00,1.00}
\definecolor{cyan2}{rgb}{0.00,0.93,0.93}
\definecolor{cyan3}{rgb}{0.00,0.80,0.80}
\definecolor{cyan4}{rgb}{0.00,0.55,0.55}
\definecolor{cyan}{rgb}{0.00,1.00,1.00}
\definecolor{darkblue}{rgb}{0.00,0.00,0.55}
\definecolor{darkcyan}{rgb}{0.00,0.55,0.55}
\definecolor{darkgoldenrod}{rgb}{0.72,0.53,0.04}
\definecolor{darkgray}{rgb}{0.66,0.66,0.66}
\definecolor{darkgreen}{rgb}{0.00,0.39,0.00}
\definecolor{darkgrey}{rgb}{0.66,0.66,0.66}
\definecolor{darkkhaki}{rgb}{0.74,0.72,0.42}
\definecolor{darkmagenta}{rgb}{0.55,0.00,0.55}
\definecolor{darkolive}{rgb}{0.33,0.42,0.18}
\definecolor{darkorange}{rgb}{1.00,0.55,0.00}
\definecolor{darkorchid}{rgb}{0.60,0.20,0.80}
\definecolor{darkred}{rgb}{0.55,0.00,0.00}
\definecolor{darksalmon}{rgb}{0.91,0.59,0.48}
\definecolor{darksea}{rgb}{0.56,0.74,0.56}
\definecolor{darkslate}{rgb}{0.18,0.31,0.31}
\definecolor{darkslate}{rgb}{0.18,0.31,0.31}
\definecolor{darkslate}{rgb}{0.28,0.24,0.55}
\definecolor{darkturquoise}{rgb}{0.00,0.81,0.82}
\definecolor{darkviolet}{rgb}{0.58,0.00,0.83}
\definecolor{deeppink}{rgb}{1.00,0.08,0.58}
\definecolor{deepsky}{rgb}{0.00,0.75,1.00}
\definecolor{dimgray}{rgb}{0.41,0.41,0.41}
\definecolor{dimgrey}{rgb}{0.41,0.41,0.41}
\definecolor{dodgerblue}{rgb}{0.12,0.56,1.00}
\definecolor{firebrick1}{rgb}{1.00,0.19,0.19}
\definecolor{firebrick2}{rgb}{0.93,0.17,0.17}
\definecolor{firebrick3}{rgb}{0.80,0.15,0.15}
\definecolor{firebrick4}{rgb}{0.55,0.10,0.10}
\definecolor{firebrick}{rgb}{0.70,0.13,0.13}
\definecolor{floralwhite}{rgb}{1.00,0.98,0.94}
\definecolor{forestgreen}{rgb}{0.13,0.55,0.13}
\definecolor{gainsboro}{rgb}{0.86,0.86,0.86}
\definecolor{ghostwhite}{rgb}{0.97,0.97,1.00}
\definecolor{gold1}{rgb}{1.00,0.84,0.00}
\definecolor{gold2}{rgb}{0.93,0.79,0.00}
\definecolor{gold3}{rgb}{0.80,0.68,0.00}
\definecolor{gold4}{rgb}{0.55,0.46,0.00}
\definecolor{goldenrod1}{rgb}{1.00,0.76,0.15}
\definecolor{goldenrod2}{rgb}{0.93,0.71,0.13}
\definecolor{goldenrod3}{rgb}{0.80,0.61,0.11}
\definecolor{goldenrod4}{rgb}{0.55,0.41,0.08}
\definecolor{goldenrod}{rgb}{0.85,0.65,0.13}
\definecolor{gold}{rgb}{1.00,0.84,0.00}
\definecolor{gray0}{rgb}{0.00,0.00,0.00}
\definecolor{gray100}{rgb}{1.00,1.00,1.00}
\definecolor{gray10}{rgb}{0.10,0.10,0.10}
\definecolor{gray11}{rgb}{0.11,0.11,0.11}
\definecolor{gray12}{rgb}{0.12,0.12,0.12}
\definecolor{gray13}{rgb}{0.13,0.13,0.13}
\definecolor{gray14}{rgb}{0.14,0.14,0.14}
\definecolor{gray15}{rgb}{0.15,0.15,0.15}
\definecolor{gray16}{rgb}{0.16,0.16,0.16}
\definecolor{gray17}{rgb}{0.17,0.17,0.17}
\definecolor{gray18}{rgb}{0.18,0.18,0.18}
\definecolor{gray19}{rgb}{0.19,0.19,0.19}
\definecolor{gray1}{rgb}{0.01,0.01,0.01}
\definecolor{gray20}{rgb}{0.20,0.20,0.20}
\definecolor{gray21}{rgb}{0.21,0.21,0.21}
\definecolor{gray22}{rgb}{0.22,0.22,0.22}
\definecolor{gray23}{rgb}{0.23,0.23,0.23}
\definecolor{gray24}{rgb}{0.24,0.24,0.24}
\definecolor{gray25}{rgb}{0.25,0.25,0.25}
\definecolor{gray26}{rgb}{0.26,0.26,0.26}
\definecolor{gray27}{rgb}{0.27,0.27,0.27}
\definecolor{gray28}{rgb}{0.28,0.28,0.28}
\definecolor{gray29}{rgb}{0.29,0.29,0.29}
\definecolor{gray2}{rgb}{0.02,0.02,0.02}
\definecolor{gray30}{rgb}{0.30,0.30,0.30}
\definecolor{gray31}{rgb}{0.31,0.31,0.31}
\definecolor{gray32}{rgb}{0.32,0.32,0.32}
\definecolor{gray33}{rgb}{0.33,0.33,0.33}
\definecolor{gray34}{rgb}{0.34,0.34,0.34}
\definecolor{gray35}{rgb}{0.35,0.35,0.35}
\definecolor{gray36}{rgb}{0.36,0.36,0.36}
\definecolor{gray37}{rgb}{0.37,0.37,0.37}
\definecolor{gray38}{rgb}{0.38,0.38,0.38}
\definecolor{gray39}{rgb}{0.39,0.39,0.39}
\definecolor{gray3}{rgb}{0.03,0.03,0.03}
\definecolor{gray40}{rgb}{0.40,0.40,0.40}
\definecolor{gray41}{rgb}{0.41,0.41,0.41}
\definecolor{gray42}{rgb}{0.42,0.42,0.42}
\definecolor{gray43}{rgb}{0.43,0.43,0.43}
\definecolor{gray44}{rgb}{0.44,0.44,0.44}
\definecolor{gray45}{rgb}{0.45,0.45,0.45}
\definecolor{gray46}{rgb}{0.46,0.46,0.46}
\definecolor{gray47}{rgb}{0.47,0.47,0.47}
\definecolor{gray48}{rgb}{0.48,0.48,0.48}
\definecolor{gray49}{rgb}{0.49,0.49,0.49}
\definecolor{gray4}{rgb}{0.04,0.04,0.04}
\definecolor{gray50}{rgb}{0.50,0.50,0.50}
\definecolor{gray51}{rgb}{0.51,0.51,0.51}
\definecolor{gray52}{rgb}{0.52,0.52,0.52}
\definecolor{gray53}{rgb}{0.53,0.53,0.53}
\definecolor{gray54}{rgb}{0.54,0.54,0.54}
\definecolor{gray55}{rgb}{0.55,0.55,0.55}
\definecolor{gray56}{rgb}{0.56,0.56,0.56}
\definecolor{gray57}{rgb}{0.57,0.57,0.57}
\definecolor{gray58}{rgb}{0.58,0.58,0.58}
\definecolor{gray59}{rgb}{0.59,0.59,0.59}
\definecolor{gray5}{rgb}{0.05,0.05,0.05}
\definecolor{gray60}{rgb}{0.60,0.60,0.60}
\definecolor{gray61}{rgb}{0.61,0.61,0.61}
\definecolor{gray62}{rgb}{0.62,0.62,0.62}
\definecolor{gray63}{rgb}{0.63,0.63,0.63}
\definecolor{gray64}{rgb}{0.64,0.64,0.64}
\definecolor{gray65}{rgb}{0.65,0.65,0.65}
\definecolor{gray66}{rgb}{0.66,0.66,0.66}
\definecolor{gray67}{rgb}{0.67,0.67,0.67}
\definecolor{gray68}{rgb}{0.68,0.68,0.68}
\definecolor{gray69}{rgb}{0.69,0.69,0.69}
\definecolor{gray6}{rgb}{0.06,0.06,0.06}
\definecolor{gray70}{rgb}{0.70,0.70,0.70}
\definecolor{gray71}{rgb}{0.71,0.71,0.71}
\definecolor{gray72}{rgb}{0.72,0.72,0.72}
\definecolor{gray73}{rgb}{0.73,0.73,0.73}
\definecolor{gray74}{rgb}{0.74,0.74,0.74}
\definecolor{gray75}{rgb}{0.75,0.75,0.75}
\definecolor{gray76}{rgb}{0.76,0.76,0.76}
\definecolor{gray77}{rgb}{0.77,0.77,0.77}
\definecolor{gray78}{rgb}{0.78,0.78,0.78}
\definecolor{gray79}{rgb}{0.79,0.79,0.79}
\definecolor{gray7}{rgb}{0.07,0.07,0.07}
\definecolor{gray80}{rgb}{0.80,0.80,0.80}
\definecolor{gray81}{rgb}{0.81,0.81,0.81}
\definecolor{gray82}{rgb}{0.82,0.82,0.82}
\definecolor{gray83}{rgb}{0.83,0.83,0.83}
\definecolor{gray84}{rgb}{0.84,0.84,0.84}
\definecolor{gray85}{rgb}{0.85,0.85,0.85}
\definecolor{gray86}{rgb}{0.86,0.86,0.86}
\definecolor{gray87}{rgb}{0.87,0.87,0.87}
\definecolor{gray88}{rgb}{0.88,0.88,0.88}
\definecolor{gray89}{rgb}{0.89,0.89,0.89}
\definecolor{gray8}{rgb}{0.08,0.08,0.08}
\definecolor{gray90}{rgb}{0.90,0.90,0.90}
\definecolor{gray91}{rgb}{0.91,0.91,0.91}
\definecolor{gray92}{rgb}{0.92,0.92,0.92}
\definecolor{gray93}{rgb}{0.93,0.93,0.93}
\definecolor{gray94}{rgb}{0.94,0.94,0.94}
\definecolor{gray95}{rgb}{0.95,0.95,0.95}
\definecolor{gray96}{rgb}{0.96,0.96,0.96}
\definecolor{gray97}{rgb}{0.97,0.97,0.97}
\definecolor{gray98}{rgb}{0.98,0.98,0.98}
\definecolor{gray99}{rgb}{0.99,0.99,0.99}
\definecolor{gray9}{rgb}{0.09,0.09,0.09}
\definecolor{gray}{rgb}{0.75,0.75,0.75}
\definecolor{green1}{rgb}{0.00,1.00,0.00}
\definecolor{green2}{rgb}{0.00,0.93,0.00}
\definecolor{green3}{rgb}{0.00,0.80,0.00}
\definecolor{green4}{rgb}{0.00,0.55,0.00}
\definecolor{greenyellow}{rgb}{0.68,1.00,0.18}
\definecolor{green}{rgb}{0.00,1.00,0.00}
\definecolor{grey0}{rgb}{0.00,0.00,0.00}
\definecolor{grey100}{rgb}{1.00,1.00,1.00}
\definecolor{grey10}{rgb}{0.10,0.10,0.10}
\definecolor{grey11}{rgb}{0.11,0.11,0.11}
\definecolor{grey12}{rgb}{0.12,0.12,0.12}
\definecolor{grey13}{rgb}{0.13,0.13,0.13}
\definecolor{grey14}{rgb}{0.14,0.14,0.14}
\definecolor{grey15}{rgb}{0.15,0.15,0.15}
\definecolor{grey16}{rgb}{0.16,0.16,0.16}
\definecolor{grey17}{rgb}{0.17,0.17,0.17}
\definecolor{grey18}{rgb}{0.18,0.18,0.18}
\definecolor{grey19}{rgb}{0.19,0.19,0.19}
\definecolor{grey1}{rgb}{0.01,0.01,0.01}
\definecolor{grey20}{rgb}{0.20,0.20,0.20}
\definecolor{grey21}{rgb}{0.21,0.21,0.21}
\definecolor{grey22}{rgb}{0.22,0.22,0.22}
\definecolor{grey23}{rgb}{0.23,0.23,0.23}
\definecolor{grey24}{rgb}{0.24,0.24,0.24}
\definecolor{grey25}{rgb}{0.25,0.25,0.25}
\definecolor{grey26}{rgb}{0.26,0.26,0.26}
\definecolor{grey27}{rgb}{0.27,0.27,0.27}
\definecolor{grey28}{rgb}{0.28,0.28,0.28}
\definecolor{grey29}{rgb}{0.29,0.29,0.29}
\definecolor{grey2}{rgb}{0.02,0.02,0.02}
\definecolor{grey30}{rgb}{0.30,0.30,0.30}
\definecolor{grey31}{rgb}{0.31,0.31,0.31}
\definecolor{grey32}{rgb}{0.32,0.32,0.32}
\definecolor{grey33}{rgb}{0.33,0.33,0.33}
\definecolor{grey34}{rgb}{0.34,0.34,0.34}
\definecolor{grey35}{rgb}{0.35,0.35,0.35}
\definecolor{grey36}{rgb}{0.36,0.36,0.36}
\definecolor{grey37}{rgb}{0.37,0.37,0.37}
\definecolor{grey38}{rgb}{0.38,0.38,0.38}
\definecolor{grey39}{rgb}{0.39,0.39,0.39}
\definecolor{grey3}{rgb}{0.03,0.03,0.03}
\definecolor{grey40}{rgb}{0.40,0.40,0.40}
\definecolor{grey41}{rgb}{0.41,0.41,0.41}
\definecolor{grey42}{rgb}{0.42,0.42,0.42}
\definecolor{grey43}{rgb}{0.43,0.43,0.43}
\definecolor{grey44}{rgb}{0.44,0.44,0.44}
\definecolor{grey45}{rgb}{0.45,0.45,0.45}
\definecolor{grey46}{rgb}{0.46,0.46,0.46}
\definecolor{grey47}{rgb}{0.47,0.47,0.47}
\definecolor{grey48}{rgb}{0.48,0.48,0.48}
\definecolor{grey49}{rgb}{0.49,0.49,0.49}
\definecolor{grey4}{rgb}{0.04,0.04,0.04}
\definecolor{grey50}{rgb}{0.50,0.50,0.50}
\definecolor{grey51}{rgb}{0.51,0.51,0.51}
\definecolor{grey52}{rgb}{0.52,0.52,0.52}
\definecolor{grey53}{rgb}{0.53,0.53,0.53}
\definecolor{grey54}{rgb}{0.54,0.54,0.54}
\definecolor{grey55}{rgb}{0.55,0.55,0.55}
\definecolor{grey56}{rgb}{0.56,0.56,0.56}
\definecolor{grey57}{rgb}{0.57,0.57,0.57}
\definecolor{grey58}{rgb}{0.58,0.58,0.58}
\definecolor{grey59}{rgb}{0.59,0.59,0.59}
\definecolor{grey5}{rgb}{0.05,0.05,0.05}
\definecolor{grey60}{rgb}{0.60,0.60,0.60}
\definecolor{grey61}{rgb}{0.61,0.61,0.61}
\definecolor{grey62}{rgb}{0.62,0.62,0.62}
\definecolor{grey63}{rgb}{0.63,0.63,0.63}
\definecolor{grey64}{rgb}{0.64,0.64,0.64}
\definecolor{grey65}{rgb}{0.65,0.65,0.65}
\definecolor{grey66}{rgb}{0.66,0.66,0.66}
\definecolor{grey67}{rgb}{0.67,0.67,0.67}
\definecolor{grey68}{rgb}{0.68,0.68,0.68}
\definecolor{grey69}{rgb}{0.69,0.69,0.69}
\definecolor{grey6}{rgb}{0.06,0.06,0.06}
\definecolor{grey70}{rgb}{0.70,0.70,0.70}
\definecolor{grey71}{rgb}{0.71,0.71,0.71}
\definecolor{grey72}{rgb}{0.72,0.72,0.72}
\definecolor{grey73}{rgb}{0.73,0.73,0.73}
\definecolor{grey74}{rgb}{0.74,0.74,0.74}
\definecolor{grey75}{rgb}{0.75,0.75,0.75}
\definecolor{grey76}{rgb}{0.76,0.76,0.76}
\definecolor{grey77}{rgb}{0.77,0.77,0.77}
\definecolor{grey78}{rgb}{0.78,0.78,0.78}
\definecolor{grey79}{rgb}{0.79,0.79,0.79}
\definecolor{grey7}{rgb}{0.07,0.07,0.07}
\definecolor{grey80}{rgb}{0.80,0.80,0.80}
\definecolor{grey81}{rgb}{0.81,0.81,0.81}
\definecolor{grey82}{rgb}{0.82,0.82,0.82}
\definecolor{grey83}{rgb}{0.83,0.83,0.83}
\definecolor{grey84}{rgb}{0.84,0.84,0.84}
\definecolor{grey85}{rgb}{0.85,0.85,0.85}
\definecolor{grey86}{rgb}{0.86,0.86,0.86}
\definecolor{grey87}{rgb}{0.87,0.87,0.87}
\definecolor{grey88}{rgb}{0.88,0.88,0.88}
\definecolor{grey89}{rgb}{0.89,0.89,0.89}
\definecolor{grey8}{rgb}{0.08,0.08,0.08}
\definecolor{grey90}{rgb}{0.90,0.90,0.90}
\definecolor{grey91}{rgb}{0.91,0.91,0.91}
\definecolor{grey92}{rgb}{0.92,0.92,0.92}
\definecolor{grey93}{rgb}{0.93,0.93,0.93}
\definecolor{grey94}{rgb}{0.94,0.94,0.94}
\definecolor{grey95}{rgb}{0.95,0.95,0.95}
\definecolor{grey96}{rgb}{0.96,0.96,0.96}
\definecolor{grey97}{rgb}{0.97,0.97,0.97}
\definecolor{grey98}{rgb}{0.98,0.98,0.98}
\definecolor{grey99}{rgb}{0.99,0.99,0.99}
\definecolor{grey9}{rgb}{0.09,0.09,0.09}
\definecolor{grey}{rgb}{0.75,0.75,0.75}
\definecolor{honeydew1}{rgb}{0.94,1.00,0.94}
\definecolor{honeydew2}{rgb}{0.88,0.93,0.88}
\definecolor{honeydew3}{rgb}{0.76,0.80,0.76}
\definecolor{honeydew4}{rgb}{0.51,0.55,0.51}
\definecolor{honeydew}{rgb}{0.94,1.00,0.94}
\definecolor{hotpink}{rgb}{1.00,0.41,0.71}
\definecolor{indianred}{rgb}{0.80,0.36,0.36}
\definecolor{ivory1}{rgb}{1.00,1.00,0.94}
\definecolor{ivory2}{rgb}{0.93,0.93,0.88}
\definecolor{ivory3}{rgb}{0.80,0.80,0.76}
\definecolor{ivory4}{rgb}{0.55,0.55,0.51}
\definecolor{ivory}{rgb}{1.00,1.00,0.94}
\definecolor{khaki1}{rgb}{1.00,0.96,0.56}
\definecolor{khaki2}{rgb}{0.93,0.90,0.52}
\definecolor{khaki3}{rgb}{0.80,0.78,0.45}
\definecolor{khaki4}{rgb}{0.55,0.53,0.31}
\definecolor{khaki}{rgb}{0.94,0.90,0.55}
\definecolor{lavenderblush}{rgb}{1.00,0.94,0.96}
\definecolor{lavender}{rgb}{0.90,0.90,0.98}
\definecolor{lawngreen}{rgb}{0.49,0.99,0.00}
\definecolor{lemonchiffon}{rgb}{1.00,0.98,0.80}
\definecolor{lightblue}{rgb}{0.68,0.85,0.90}
\definecolor{lightcoral}{rgb}{0.94,0.50,0.50}
\definecolor{lightcyan}{rgb}{0.88,1.00,1.00}
\definecolor{lightgoldenrod}{rgb}{0.93,0.87,0.51}
\definecolor{lightgoldenrod}{rgb}{0.98,0.98,0.82}
\definecolor{lightgray}{rgb}{0.83,0.83,0.83}
\definecolor{lightgreen}{rgb}{0.56,0.93,0.56}
\definecolor{lightgrey}{rgb}{0.83,0.83,0.83}
\definecolor{lightpink}{rgb}{1.00,0.71,0.76}
\definecolor{lightsalmon}{rgb}{1.00,0.63,0.48}
\definecolor{lightsea}{rgb}{0.13,0.70,0.67}
\definecolor{lightsky}{rgb}{0.53,0.81,0.98}
\definecolor{lightslate}{rgb}{0.47,0.53,0.60}
\definecolor{lightslate}{rgb}{0.47,0.53,0.60}
\definecolor{lightslate}{rgb}{0.52,0.44,1.00}
\definecolor{lightsteel}{rgb}{0.69,0.77,0.87}
\definecolor{lightyellow}{rgb}{1.00,1.00,0.88}
\definecolor{limegreen}{rgb}{0.20,0.80,0.20}
\definecolor{linen}{rgb}{0.98,0.94,0.90}
\definecolor{magenta1}{rgb}{1.00,0.00,1.00}
\definecolor{magenta2}{rgb}{0.93,0.00,0.93}
\definecolor{magenta3}{rgb}{0.80,0.00,0.80}
\definecolor{magenta4}{rgb}{0.55,0.00,0.55}
\definecolor{magenta}{rgb}{1.00,0.00,1.00}
\definecolor{maroon1}{rgb}{1.00,0.20,0.70}
\definecolor{maroon2}{rgb}{0.93,0.19,0.65}
\definecolor{maroon3}{rgb}{0.80,0.16,0.56}
\definecolor{maroon4}{rgb}{0.55,0.11,0.38}
\definecolor{maroon}{rgb}{0.69,0.19,0.38}
\definecolor{mediumaquamarine}{rgb}{0.40,0.80,0.67}
\definecolor{mediumblue}{rgb}{0.00,0.00,0.80}
\definecolor{mediumorchid}{rgb}{0.73,0.33,0.83}
\definecolor{mediumpurple}{rgb}{0.58,0.44,0.86}
\definecolor{mediumsea}{rgb}{0.24,0.70,0.44}
\definecolor{mediumslate}{rgb}{0.48,0.41,0.93}
\definecolor{mediumspring}{rgb}{0.00,0.98,0.60}
\definecolor{mediumturquoise}{rgb}{0.28,0.82,0.80}
\definecolor{mediumviolet}{rgb}{0.78,0.08,0.52}
\definecolor{midnightblue}{rgb}{0.10,0.10,0.44}
\definecolor{mintcream}{rgb}{0.96,1.00,0.98}
\definecolor{mistyrose}{rgb}{1.00,0.89,0.88}
\definecolor{moccasin}{rgb}{1.00,0.89,0.71}
\definecolor{navajowhite}{rgb}{1.00,0.87,0.68}
\definecolor{navyblue}{rgb}{0.00,0.00,0.50}
\definecolor{navy}{rgb}{0.00,0.00,0.50}
\definecolor{oldlace}{rgb}{0.99,0.96,0.90}
\definecolor{olivedrab}{rgb}{0.42,0.56,0.14}
\definecolor{orange1}{rgb}{1.00,0.65,0.00}
\definecolor{orange2}{rgb}{0.93,0.60,0.00}
\definecolor{orange3}{rgb}{0.80,0.52,0.00}
\definecolor{orange4}{rgb}{0.55,0.35,0.00}
\definecolor{orangered}{rgb}{1.00,0.27,0.00}
\definecolor{orange}{rgb}{1.00,0.65,0.00}
\definecolor{orchid1}{rgb}{1.00,0.51,0.98}
\definecolor{orchid2}{rgb}{0.93,0.48,0.91}
\definecolor{orchid3}{rgb}{0.80,0.41,0.79}
\definecolor{orchid4}{rgb}{0.55,0.28,0.54}
\definecolor{orchid}{rgb}{0.85,0.44,0.84}
\definecolor{palegoldenrod}{rgb}{0.93,0.91,0.67}
\definecolor{palegreen}{rgb}{0.60,0.98,0.60}
\definecolor{paleturquoise}{rgb}{0.69,0.93,0.93}
\definecolor{paleviolet}{rgb}{0.86,0.44,0.58}
\definecolor{papayawhip}{rgb}{1.00,0.94,0.84}
\definecolor{peachpuff}{rgb}{1.00,0.85,0.73}
\definecolor{peru}{rgb}{0.80,0.52,0.25}
\definecolor{pink1}{rgb}{1.00,0.71,0.77}
\definecolor{pink2}{rgb}{0.93,0.66,0.72}
\definecolor{pink3}{rgb}{0.80,0.57,0.62}
\definecolor{pink4}{rgb}{0.55,0.39,0.42}
\definecolor{pink}{rgb}{1.00,0.75,0.80}
\definecolor{plum1}{rgb}{1.00,0.73,1.00}
\definecolor{plum2}{rgb}{0.93,0.68,0.93}
\definecolor{plum3}{rgb}{0.80,0.59,0.80}
\definecolor{plum4}{rgb}{0.55,0.40,0.55}
\definecolor{plum}{rgb}{0.87,0.63,0.87}
\definecolor{powderblue}{rgb}{0.69,0.88,0.90}
\definecolor{purple1}{rgb}{0.61,0.19,1.00}
\definecolor{purple2}{rgb}{0.57,0.17,0.93}
\definecolor{purple3}{rgb}{0.49,0.15,0.80}
\definecolor{purple4}{rgb}{0.33,0.10,0.55}
\definecolor{purple}{rgb}{0.63,0.13,0.94}
\definecolor{red1}{rgb}{1.00,0.00,0.00}
\definecolor{red2}{rgb}{0.93,0.00,0.00}
\definecolor{red3}{rgb}{0.80,0.00,0.00}
\definecolor{red4}{rgb}{0.55,0.00,0.00}
\definecolor{red}{rgb}{1.00,0.00,0.00}
\definecolor{rosybrown}{rgb}{0.74,0.56,0.56}
\definecolor{royalblue}{rgb}{0.25,0.41,0.88}
\definecolor{saddlebrown}{rgb}{0.55,0.27,0.07}
\definecolor{salmon1}{rgb}{1.00,0.55,0.41}
\definecolor{salmon2}{rgb}{0.93,0.51,0.38}
\definecolor{salmon3}{rgb}{0.80,0.44,0.33}
\definecolor{salmon4}{rgb}{0.55,0.30,0.22}
\definecolor{salmon}{rgb}{0.98,0.50,0.45}
\definecolor{sandybrown}{rgb}{0.96,0.64,0.38}
\definecolor{seagreen}{rgb}{0.18,0.55,0.34}
\definecolor{seashell1}{rgb}{1.00,0.96,0.93}
\definecolor{seashell2}{rgb}{0.93,0.90,0.87}
\definecolor{seashell3}{rgb}{0.80,0.77,0.75}
\definecolor{seashell4}{rgb}{0.55,0.53,0.51}
\definecolor{seashell}{rgb}{1.00,0.96,0.93}
\definecolor{sienna1}{rgb}{1.00,0.51,0.28}
\definecolor{sienna2}{rgb}{0.93,0.47,0.26}
\definecolor{sienna3}{rgb}{0.80,0.41,0.22}
\definecolor{sienna4}{rgb}{0.55,0.28,0.15}
\definecolor{sienna}{rgb}{0.63,0.32,0.18}
\definecolor{skyblue}{rgb}{0.53,0.81,0.92}
\definecolor{slateblue}{rgb}{0.42,0.35,0.80}
\definecolor{slategray}{rgb}{0.44,0.50,0.56}
\definecolor{slategrey}{rgb}{0.44,0.50,0.56}
\definecolor{snow1}{rgb}{1.00,0.98,0.98}
\definecolor{snow2}{rgb}{0.93,0.91,0.91}
\definecolor{snow3}{rgb}{0.80,0.79,0.79}
\definecolor{snow4}{rgb}{0.55,0.54,0.54}
\definecolor{snow}{rgb}{1.00,0.98,0.98}
\definecolor{springgreen}{rgb}{0.00,1.00,0.50}
\definecolor{steelblue}{rgb}{0.27,0.51,0.71}
\definecolor{tan1}{rgb}{1.00,0.65,0.31}
\definecolor{tan2}{rgb}{0.93,0.60,0.29}
\definecolor{tan3}{rgb}{0.80,0.52,0.25}
\definecolor{tan4}{rgb}{0.55,0.35,0.17}
\definecolor{tan}{rgb}{0.82,0.71,0.55}
\definecolor{thistle1}{rgb}{1.00,0.88,1.00}
\definecolor{thistle2}{rgb}{0.93,0.82,0.93}
\definecolor{thistle3}{rgb}{0.80,0.71,0.80}
\definecolor{thistle4}{rgb}{0.55,0.48,0.55}
\definecolor{thistle}{rgb}{0.85,0.75,0.85}
\definecolor{tomato1}{rgb}{1.00,0.39,0.28}
\definecolor{tomato2}{rgb}{0.93,0.36,0.26}
\definecolor{tomato3}{rgb}{0.80,0.31,0.22}
\definecolor{tomato4}{rgb}{0.55,0.21,0.15}
\definecolor{tomato}{rgb}{1.00,0.39,0.28}
\definecolor{turquoise1}{rgb}{0.00,0.96,1.00}
\definecolor{turquoise2}{rgb}{0.00,0.90,0.93}
\definecolor{turquoise3}{rgb}{0.00,0.77,0.80}
\definecolor{turquoise4}{rgb}{0.00,0.53,0.55}
\definecolor{turquoise}{rgb}{0.25,0.88,0.82}
\definecolor{violetred}{rgb}{0.82,0.13,0.56}
\definecolor{violet}{rgb}{0.93,0.51,0.93}
\definecolor{wheat1}{rgb}{1.00,0.91,0.73}
\definecolor{wheat2}{rgb}{0.93,0.85,0.68}
\definecolor{wheat3}{rgb}{0.80,0.73,0.59}
\definecolor{wheat4}{rgb}{0.55,0.49,0.40}
\definecolor{wheat}{rgb}{0.96,0.87,0.70}
\definecolor{whitesmoke}{rgb}{0.96,0.96,0.96}
\definecolor{white}{rgb}{1.00,1.00,1.00}
\definecolor{yellow1}{rgb}{1.00,1.00,0.00}
\definecolor{yellow2}{rgb}{0.93,0.93,0.00}
\definecolor{yellow3}{rgb}{0.80,0.80,0.00}
\definecolor{yellow4}{rgb}{0.55,0.55,0.00}
\definecolor{yellowgreen}{rgb}{0.60,0.80,0.20}
\definecolor{yellow}{rgb}{1.00,1.00,0.00}

\definecolor{unsure1}{rgb}{1.00,0.00,0.46}
\definecolor{colorFrontal}{rgb}{0.93,0.27,0.21}
\definecolor{colorLateral}{rgb}{0.68,0.66,0.80}
\definecolor{colorHorizontal}{rgb}{0.94,0.63,0.37}

\definecolor{myorange}{rgb}{1.00,0.50,0.25}     
\definecolor{mypink}{rgb}{1.00,0.00,1.00}       
\definecolor{mybluesea}{rgb}{0.00,1.00,1.00}    
\definecolor{mycobalt}{rgb}{0.40,0.40,1.00}     
\definecolor{myyellow}{rgb}{1.00,1.00,0.00}     
\definecolor{mygray}{rgb}{0.50,0.50,0.50}       

\newcommand{\mat}[1]{\mathbf{#1}}                       
\newcommand{\ten}[1]{\mathbf{\underline{#1}}}           
\newcommand{\hmat}[1]{\mathbf{\hat{#1}}}                
\newcommand{\cmat}[1]{\mathbf{\bar{#1}}}                
\newcommand{\imat}[1]{\mathbf{\breve{#1}}}              
\newcommand{\frob}[1]{\left \|  #1 \right \|_F}
\newcommand{\cpd}[1]{\llbracket #1 \rrbracket}
\newcommand{\R}{\mathbb{R}}
\tcbuselibrary{skins}
\newcolumntype{Y}{>{\raggedleft\arraybackslash}X}
\tcbset{tab2/.style={enhanced,
                    fonttitle=\bfseries,
                    fontupper=\normalsize\sffamily,
                    colback=yellow!10!white,
                    colframe=red!50!black,
                    colbacktitle=Salmon!40!white,
                    coltitle=black,center title
                    }
}

\newcommand{\includeLLfigure}{
    \begin{figure}[t]
        \includegraphics[width=\linewidth]{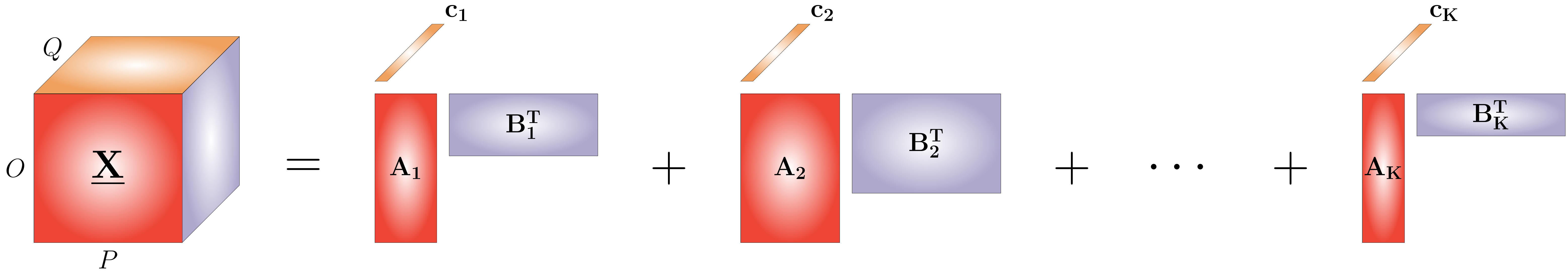}
        \caption{The LL1 decomposition is a combination of the CPD and the HOSVD. The tensor $\ten{X}_k = (\mat{A}_k \mat{B}^T_k) \circ \mat{c}_k$ still exhibits the simplest structure within the LL1 decomposition framework.}
        \label{fig:ll1}
    \end{figure}
}

\newcommand{\includeBambyFigure}{
    \begin{figure}[t]
        \includegraphics[width=\linewidth]{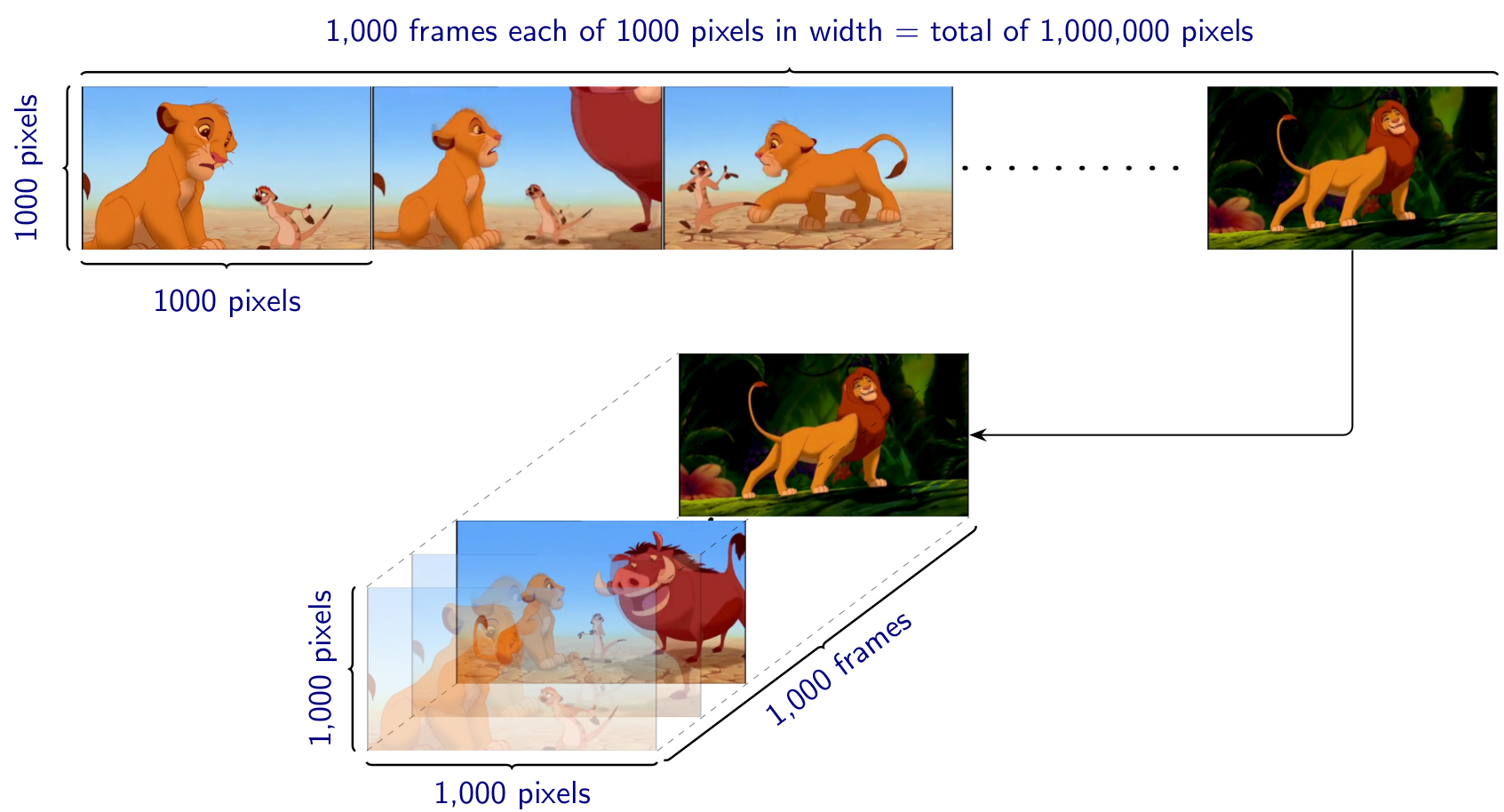}
        \caption{Efficient representation of an imbalanced block-matrix structure (a set of video frames, top row) in the form of much more convenient and flexible tensor structure (a cube of frames, bottom row).}
        \label{fig:tensor_bamby}
    \end{figure}
}

\newcommand{\includeCPDFigure}{
    \begin{figure}[t]
        \includegraphics[width=\linewidth]{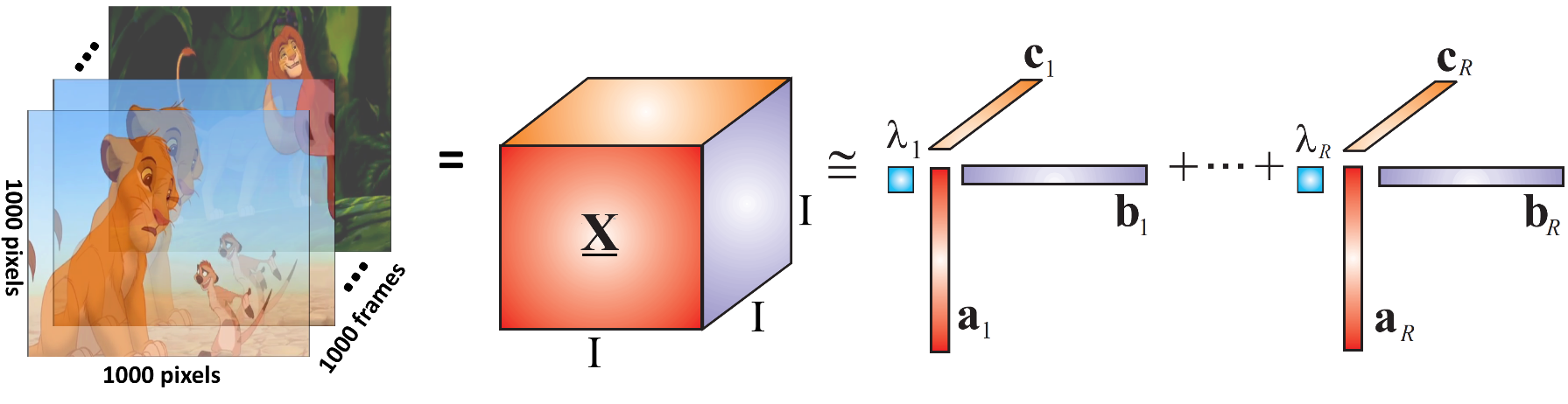}
        \caption{The Canonical Polyadic Decomposition (CPD). The tensor, $\ten{X}$, is represented as a sum of rank-1 tensors, $\ten{X}_r$, given by the outer product of the factor vectors $\mat{a}_r, \mat{b}_r, \mat{c}_r$.}
        \label{fig:cpd_bamby}
    \end{figure}
}

\newcommand{\includeFaceTensorFigure}{
    \begin{figure}[t]
        \includegraphics[width=0.45\linewidth]{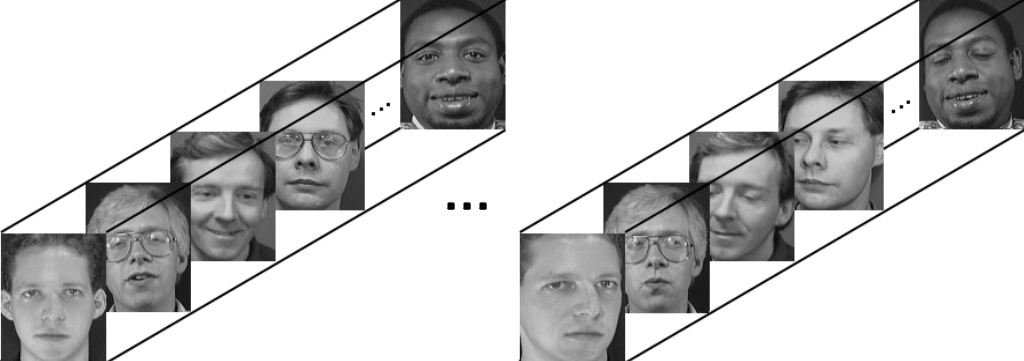}
        \hfill
        {\vrule width 1.5pt}
        \hfill
        \includegraphics[width=0.5\linewidth]{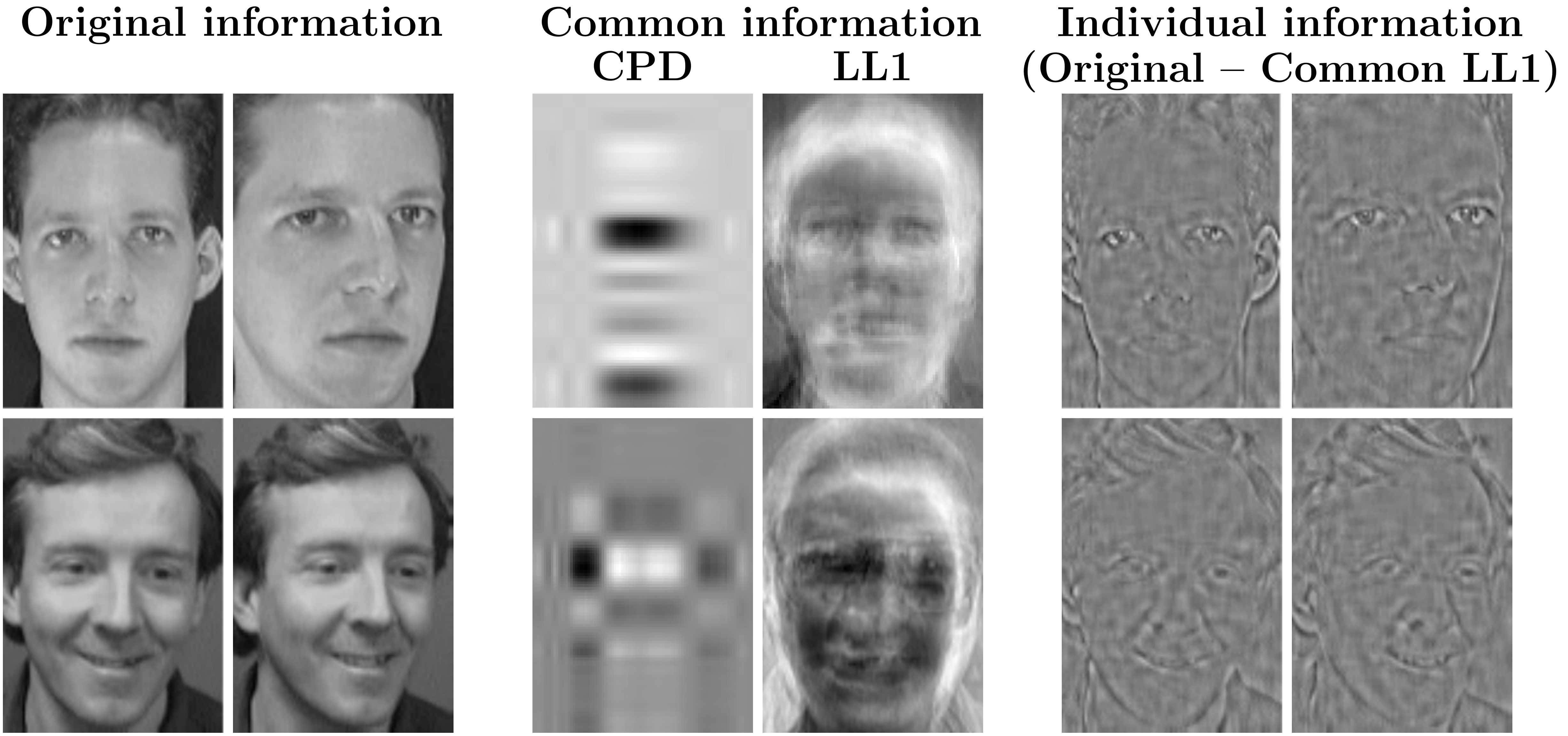}
        \caption{\textbf{Left column:} Examples of the tensor representations for $\ten{X}^i \in \R^{112 \times 92 \times 40}$ where the images of 40 different subjects were stacked along the third dimension during the training stage for tensor-based common and individual feature extraction.
        \textbf{Right column:} Common and individual feature extraction. Left: Examples of images in the ORL dataset. Center: Examples of common features computed through the CPD and LL1 decompositions. Right: Examples of extracted individual features obtained by subtraction of the common information from the corresponding original images.}
        \label{fig:train_sample}
        \label{fig:res_orig_indiv_com}
    \end{figure}
}

\newcommand{\includeColorEnsemble}{
    \begin{figure}[t]
        \includegraphics[width=\linewidth]{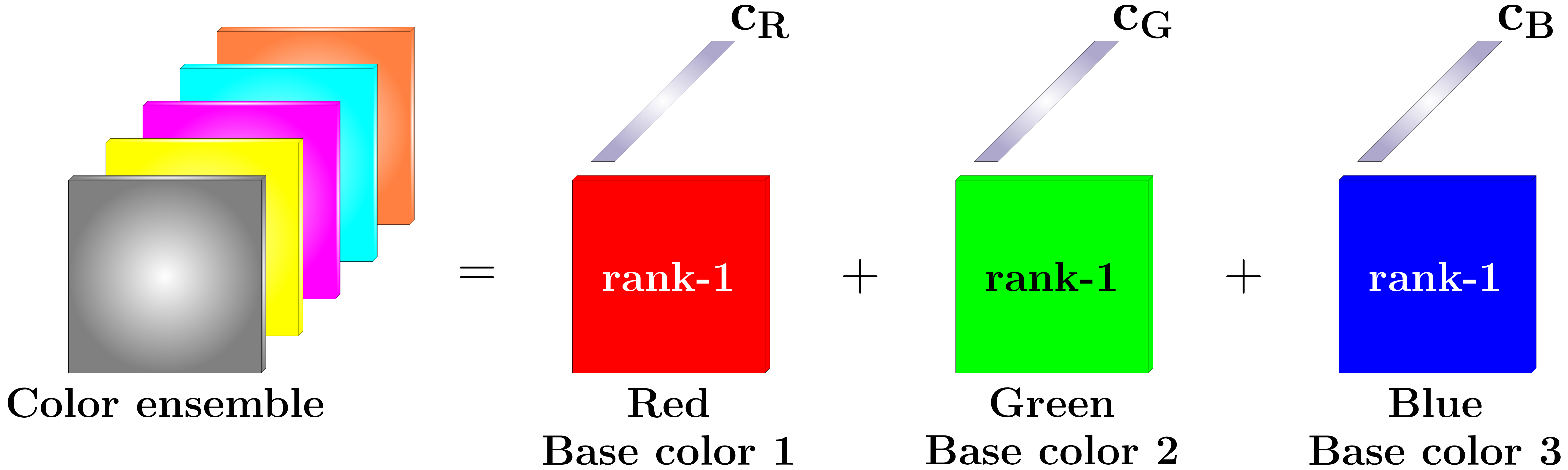}
        \caption{Link between the outer product and common features for multidimensional data. The ensemble consists of the grey, yellow, magenta, bright blue and brown-orange colors each of which is a combination of three base colors: red, green, blue.}
        \label{fig:color_ensemble}
    \end{figure}
}

\title{Tensor Valued Common and Individual Feature Extraction: Multi-dimensional Perspective}
\author[1]{Ilia Kisil}
\author[1]{Giuseppe G. Calvi}
\author[1]{Danilo P. Mandic}
\affil[1]{\small
    Electrical and Electronic Engineering Department, Imperial College London, SW7 2AZ, UK\\
    E-mails: \{i.kisil15, giuseppe.calvi15, d.mandic\}@imperial.ac.uk
}
\hypersetup{
    pdfpagelayout={},
    pdftitle={Tensor Valued Common and Individual Feature Extraction: Multi-dimensional Perspective},
    pdfsubject={},
    pdfauthor={Ilia Kisil, Giuseppe G. Calvi, Danilo P. Mandic},
    pdfkeywords={Tensor decomposition, tensor rank, feature extraction, common and individual features, classification},
    pdfstartview=FitH,
    pdfpagemode={UseOutlines},
    bookmarksnumbered=true, bookmarksopen=true, colorlinks,
    citecolor=black,%
    filecolor=black,%
    linkcolor=black,%
    urlcolor=black
}

\begin{document}

    \maketitle

    \begin{abstract}
        A novel method for common and individual feature analysis from exceedingly large-scale data is proposed, in order to ensure the tractability of both the computation and storage and thus mitigate the curse of dimensionality, a major bottleneck in modern data science. This is achieved by making use of the inherent redundancy in so-called multi-block data structures, which represent multiple observations of the same phenomenon taken at different times, angles or recording conditions. Upon providing an intrinsic link between the properties of the outer vector product and extracted features in tensor decompositions (TDs), the proposed common and individual information extraction from multi-block data is performed through imposing physical meaning to otherwise unconstrained factorisation approaches. This is shown to  dramatically reduce the dimensionality of search spaces for subsequent classification procedures and to yield greatly enhanced accuracy. Simulations on a multi-class classification task of large-scale extraction of individual features from a collection of partially related real-world images demonstrate the advantages of the ``blessing of dimensionality'' associated with TDs.
    \end{abstract}

    \providecommand{\keywords}[1]{\textbf{\textit{Index terms---}} #1}

    \section{Introduction}
        Modern datasets in data science applications have immense volume, veracity, velocity and variety (the for V`s of big data) \cite{cichocki2014era,cichocki2016tensor}, and often exhibit a large degree of structural richness among their entries. These data characteristics are often prohibitive to the application of classical matrix algebra as its ``flat-view'' way of operation cannot cope with the sheer volume of data and the corresponding imbalanced matrix structures, such as as ``tall and narrow'' or ``short and wide'' ones. On the other hand, when arranged in multi-dimensional structures (tensors), the same data often admit much more convenient and mathematically tractable ways of analysis, by virtue of the associated multi-linear algebra.
However, until recently, such an approach to data analysis was not very popular, due to high demand for storage and computational resources.

There are several ways to \textit{tensorize} data prior to further analysis, such as through: (i) natural tensor formation, (ii) experimental design, or (iii) mathematical construction \cite{cichocki2015tensor}. This flexibility and a highly informative nature of multi-way data representation is supported by tensor decompositions (TDs) which allow for storage and memory efficient low-rank approximation of otherwise intractable large data, and are being exploited in diverse range disciplines including
brain science \cite{cichocki2013tensor,escudero2015multiscale},
chemometrics \cite{smilde2005multi},
psychometric \cite{kiers2001three},
machine learning \cite{sidiropoulos2016tensor,nguyen2015tensor} and
signal processing \cite{cichocki2015tensor}.

The generalisation of a matrix to a tensor, as in \cref{fig:tensor_bamby}, is intuitive but highly non-trivial, not least due to multi-linear algebra having different properties to linear algebra. Along these lines, the authors in \cite{cong2015tensor} consider the physical meaning of factor matrices obtained through TDs. Missing data can also be handled through tensor dictionary learning \cite{phan2013tensor}, whereby the tensor structure allows for a simultaneous retrieval of local patterns and establishing the global information.
The algorithms for classification of the multi-dimensional data have been proposed in \cite{savas2007handwritten,zink2016tensor}.

We here consider a problem of the extraction of information and  classification of reduced dimension features from large-scale multi-block data.
A typical example of such structure is a set of recordings of the same phenomenon but under different experimental setups, such as multiple images of objects recorded under different lighting and angle combination, multi-block data. Intuitively, the so obtained ensemble of images contains some common and some individual features, and machine learning tasks would benefit from exploiting only either common features (for clustering) or individual features (for identification), both of much lower dimensionality than the original data.
\includeBambyFigure
In this work, to resolve the computational and storage issues for large scale
classification problems, the identification of common features is achieved by
first providing an additional insight into the physical meaning of the outer product of multiple
vectors in the tensor setting, supported by an intuitive example. The separation of the
common and individual feature subspaces is then achieved by multi-linear
rank decomposition (LL1), whereby the number of ``simplest''  data
structures in such a decomposition is equivalent to the number of multi-linear
tensor ranks \cite{sorber2013optimization}. The non-negativity constraint is further imposed on the
so extracted factors, to preserve the physical properties of the images
considered. Simulations on the benchmark ORL dataset demonstrate that
the proposed method provides significant advantages in terms of accuracy, mathematical tractability and ease of interpretation, when used in
conjunction with standard classification algorithms.

    \section{Common and Individual Components in Data} \label{sec:relatedwork}
        
Consider a set, $\mathcal{X}$, of $N$ observations in a matrix form, given by
\begin{equation}
    \mathcal{X} = \{\mat{X}_n \in \R^{I \times J_n} : n = 1,2,\ldots, N\}
    \label{eq:matrix_set}
\end{equation}
where the so called block-matrix structure $\mathcal{X}$ could be a representation of medical images, EEG recordings, or financial stock characteristics. All members of such a set of matrices $\mathcal{X}$ are naturally linked together and it would be beneficial to analyse them simultaneously at the same time, however, the representation in \eqref{eq:matrix_set} yields imbalanced (tall and narrow) structure which is cumbersome for further processing. The main goal of the common and individual feature analysis is, therefore, to make use of the ``blessing of dimensionality'' associated with tensor structures, in order to find a much lower-dimensional unique subspace $\cmat{A}$ that is common across all $n \in N$. In this way, the common subspace, $\cmat{A}$, can be separated from the individual information, $\imat{A}_n$, for every $n$.

The flat view matrix methods typically stack all entries of $\mathcal{X}$ into a tall and narrow matrix $\mat{X} = \big[ \mat{X}_1^T, \mat{X}_2^T, \dots , \mat{X}_N^T\big]^T$ and subsequently perform matrix factorisations, such as the principal component analysis (PCA) \cite{guo2008unified}, to give:
\begin{equation}
    \mat{X} = \imat{A}\cmat{A}^T
\end{equation}
where $\imat{A} = \big[ \imat{A}_1^T, \imat{A}_2^T, \dots , \imat{A}_N^T \big]^T$. In \cite{groves2011linked}, this method was applied to neuroimaging data of patients with Alzheimer disease, whereby $\cmat{A}$ is interpreted as well established knowledge about the disease (common components), while $\imat{A}_n$ represents the individual state for a specific patient. However, as with all matrix models, this approach does not generalise well and is only appropriate when all components, $\mat{X}_n$, of the tall and narrow matrix, $\mat{X}$, exhibit exactly the same common information.

Approaches to common and individual feature extraction presented in \cite{lock2013joint,zhou2016group} also employ a PCA like factorisation to every entry of the naturally linked dataset given in \eqref{eq:matrix_set}, to yield
\begin{equation}
    \begin{aligned}
        \mat{X}_n = \mat{A}_n\mat{B}_n^T &= \begin{bmatrix} \cmat{A} & \imat{A}_n \end{bmatrix} \begin{bmatrix} \cmat{B}_n^T \\[0.3em] \imat{B}_n^T \end{bmatrix}\\
        &= \cmat{A}\cmat{B}_n^T + \imat{A}_n\imat{B}_n^T = \cmat{X}_n + \imat{X}_n
        \label{eq:com_indiv_comb}
    \end{aligned}
\end{equation}
where the matrices, $\cmat{X}_n$, are the common components across the dataset $\mathcal{X}$, while the matrices, $\imat{X}_n$, are the individual components for every $\mat{X}_n$ in $\mathcal{X}$. The matrices $\cmat{A}$ and $\imat{A}_n$ are the basis matrices respectively for the matrices $\cmat{X}_n$ and $\imat{X}_n$, while the matrices, $\cmat{B}_n$ and $\imat{B}_n$ represent mixing coefficients so that
$\cmat{X}_n = \cmat{A}\cmat{B}_n^T$ and $\imat{X}_n = \imat{A}_n\imat{B}_n^T$

\noindent \textbf{Remark 1.} Due to the linear separability of the matrices $\cmat{X}_n$ and $\imat{X}_n$, it is sufficient to establish the basis of the common information, $\cmat{A}$, which can be estimated through iterative minimisation of the cost function, formulated in \cite{zhou2015common} as:
\begin{equation}
    J(\mat{Q}_n, \mat{z}_{(n,m)}, \mat{a}_m) = \sum_{n=1}^{N}\frob{\mat{Q}_n\mat{z}_{(n,m)} - \mat{a}_m}^2
    \label{eq:cife_cost}
\end{equation}
where the orthogonal matrix $\mat{Q}_n$ is obtained from $\mat{X}_n = \mat{Q}_n\mat{R}_n$, $\mat{z}_{(n,m)}$ is the column vector of $\mat{Z}_n = \mat{R}_n (\mat{B}_n^T)^\dagger$ and $\mat{a}_m$ is the common component which defines the basis of $\cmat{A}$, if the cost in \eqref{eq:cife_cost} is smaller then a predefined threshold.
Thus, the weak but consistently presented similarities among data matrices from the dataset $\mathcal{X}$ contribute to the total cost in \eqref{eq:cife_cost} the same amount as the very prominent ones and, therefore, take important part in the multi-block data analysis.

The matrix approaches have their pros and cons and are powerful if exploited appropriately, however, they do not account directly for the intrinsic multidimensional form of data. To this end, we propose a novel method for common and individual feature extraction which exploits multi-modal properties of tensor decompositions.

        \includeCPDFigure

    \section{Notation and Theoretical Background} \label{sec:background}

A tensor of \textit{order} $N$ is a N-dimensional array and is denoted by a bold underlined capital letter, $\ten{X} \in \R^{I_1 \times I_2 \times \cdots \times I_N}$. A particular dimension of $\ten{X}$ is usually referred to as a \textit{mode}. An element of a tensor is a scalar $x_{i_1,i_2,\dots,i_N} = \ten{X}(i_1,i_2,\dots,i_N)$ which has $N$ indices. A \textit{fiber} is a vector obtained by fixing all but one of the indices, e.g. $\ten{X}_{(i_1,:,i_3,\dots,i_N)}$ is the mode-2 fiber.
Fixing all but two of the indices yields a matrix called a \textit{slice} of a tensor, e.g. $\ten{X}_{(:,:,i_3,\dots,i_N)}$ is the \textit{frontal} slice. Mode-n unfolding is the process of element mapping from a tensor to a matrix, e.g. $\ten{X} \rightarrow \mat{X}_{(2)} $ is the mode-2 unfolding. A mode-n product of a tensor with a matrix is equivalent to
\begin{equation}
    \ten{Y} = \ten{X} \times_n \mat{A} \quad \Leftrightarrow \quad \mat{Y}_{(n)} = \mat{A}\mat{X}_{(n)}
\end{equation}
The outer product of $N$ vectors results in a \textit{rank-1} tensor of order $N$, e.g. $\mat{a}_1 \circ \mat{a}_2 \circ \cdots \circ \mat{a}_n = \ten{X} \in \R^{I_1 \times I_2 \times \cdots \times I_N}$

\subsection{Basic Tensor Decompositions}
The Canonical Polyadic Decomposition (CPD), illustrated in \cref{fig:cpd_bamby}, represents a given tensor $\ten{X}$ as a sum of rank-1 tensors $\ten{X}_r, r=1,2,\ldots,R$. For a third order tensor of rank $R$, the CPD is given by
\begin{equation}
    \begin{aligned}
    \ten{X} &\cong  \sum_{r=1}^R \ten{X}_r \cong  \sum_{r=1}^R \lambda_{r} \cdot \mat{a}_r \circ \mat{b}_r \circ \mat{c}_r\\
            &\cong  \ten{\Lambda} \times_1 \mat{A} \times_2 \mat{B} \times_3 \mat{C} = \cpd{\ten{\Lambda};\mat{A} , \mat{B}, \mat{C}}
    \end{aligned}
    \label{eq:cpd}
\end{equation}
where $\ten{\Lambda}$ is a superdiagonal core tensor that guarantees ``one to one relation'' for the factor vectors $\mat{a}_r,\mat{b}_r$ and $\mat{c}_r$, while $\mat{A}, \mat{B}$ and $\mat{C}$ are factor matrices which are composed of the corresponding factor vectors, e.g. $\mat{A} = \big[ \mat{a}_1, \mat{a}_2,\ldots, \mat{a}_R\big]$. Despite soft uniqueness conditions, in practice the CPD in \eqref{eq:cpd} does not provide the exact decomposition of the original data tensor  \cite{kolda2009tensor}.
On the other hand, the Higher Order Singular Value Decomposition (HOSVD) requires orthogonality constraints to be imposed on the factor matrices, is always exact \cite{de2000multilinear}, and takes the form
\begin{equation}
    \begin{aligned}
        \ten{X} &= \sum_{r_a=1}^{R_a} \sum_{r_b=1}^{R_b} \sum_{r_c=1}^{R_c} g_{r_ar_br_c} \cdot \mat{a}_{r_a} \circ \mat{b}_{r_b} \circ \mat{c}_{r_c}\\
                &= \ten{G} \times_1 \mat{A} \times_2 \mat{B} \times_3 \mat{C} = \cpd{\ten{G};\mat{A} , \mat{B}, \mat{C}}
    \end{aligned}
    \label{eq:hosvd}
\end{equation}
where $\ten{G}$ is a dense core tensor, $\mat{A}, \mat{B},$ and $\mat{C}$ are the orthogonal factor matrices and the n-tuple $(R_a, R_b, R_b)$ is called the \textit{multi-linear rank}. Observe that the HOSVD also decomposes multi-dimensional data into a sum of rank-1 terms $\mat{a}_{r_a} \circ \mat{b}_{r_b} \circ \mat{c}_{r_c}$. However, as opposed  to the ``one to one'' relation for the CPD, the HOSVD models all possible combinations of its factor vectors, hence, providing enhanced flexibility.
To make use of the desirable properties of both CPD and HOSVD, the LL1 decomposition efficiently combines their concepts \cite{de2008decompositions}, by decomposing the tensor $\ten{X}$ into a linear combination of $K$ tensors, whereby each term $\ten{X}_k = \cpd{\ten{\Lambda}_k; \mat{A}_k, \mat{B}_k, \mat{c}_k}$ has a multi-linear rank $(L_k, L_k, 1)$, that is
\begin{equation}
    \begin{aligned}
        \ten{X}
        &= \sum_{k=1}^K \ten{X}_k = \sum_{k=1}^K (\mat{A}_k \mat{B}^T_k) \circ \mat{c}_k = \sum_{k=1}^K \mat{G}_k \circ \mat{c}_k\\
        &= \sum_{k=1}^K \ten{\Lambda}_k \times_1 \mat{A}_k \times_2 \mat{B}_k \times_3 \mat{c}_k = \sum_{k=1}^K \ten{G}_k \times_3 \mat{c}_k
    \end{aligned}
    \label{eq:ll1}
\end{equation}
The LL1 decomposition is illustrated in \cref{fig:ll1}, where $\ten{X} \in \R^{O \times P \times Q}$, and  the ``one to one'' relation between the factor matrices $\mat{A}_k \in \R^{O \times L_k}$, $\mat{B}_k \in \R^{P \times L_k}$, and factor vector $\mat{c}_k \in \R^{Q}$ is preserved.
Moreover, upon employing the matrix-tensor duality, we can represent the matrix $\mat{G}_k \in \R^{O \times P}$ as a tensor of order three, $\hat{\ten{G}}_k \in \R^{O \times P \times 1}$, so that $\hat{\ten{G}}_k = \ten{G}_k$.

\noindent\textbf{Remark 2.} The matrix $\mat{G}_k$ in \eqref{eq:ll1} is no longer of rank-1 and is consequently more informative. However, the so-obtained tensor $\ten{X}_k = \mat{G}_k \circ \mat{c}_k$ is still considered to exhibit the simplest structure as far as the LL1 decomposition is concerned.

        \includeLLfigure

    \section{Common and Individual Feature Extraction} \label{sec:method}
        \newcommand{\includeProposedALG}{
\captionsetup[algorithm]{labelsep=period}
    \begin{algorithm}[t]
        \begin{algorithmic}[1]
            \REQUIRE $\ten{X} \in \R^{O \times P \times Q}$ and  K sets of multilinear tensor rank $(L_k,L_k,1)$
            \ENSURE Factor matrices $\mat{A}_k \in \R^{O \times L_k}, \mat{B}_k \in \R^{P \times L_k}, \mat{c}_k \in \R^{Q}$, and scaling vectors $\mat{\lambda}_k \in \R^{L_k}$
            \STATE Initialize factor matrices $\mat{A}_k, \mat{B}_k, \mat{c}_k$
            \WHILE {not converged or iteration limit is not reached}
                \FOR {k = 1, \dots , K}
                    \STATE $ \ten{X} = \ten{X} - \sum_{n=1, n \neq k}^N \cpd{ \ten{\Lambda}_n ; \mat{A}_n, \mat{B}_n, \mat{c}_n }$
                    \STATE $\mat{C}_k = $ repeat$(\mat{c}_k, L_k)$
                    \STATE $\hmat{A}_k = \mat{X}_{(1)}(\mat{C}_k \odot  \mat{B}_k)(\mat{C}_k^T\mat{C}_k \ast \mat{B}_k^T\mat{B}_k)^\dagger$
                    \STATE $\hmat{B}_k = \mat{X}_{(2)}(\mat{C}_k \odot  \mat{A}_k)(\mat{C}_k^T\mat{C}_k \ast \mat{A}_k^T\mat{A}_k)^\dagger$
                    \STATE $\hmat{c}_k =$ NonNegLeastSq $(\mat{A} \odot \mat{B},\mat{X}_{(3)})$
                    \STATE Normalize each column of $\hmat{A}_k,\hmat{B}_k$ and $\hmat{c}_k$ to unit length and store the norms in $\mat{\lambda}_k$
                    \STATE Assign $\hmat{A}_k \rightarrow \mat{A}_k$; $\hmat{B}_k \rightarrow \mat{B}_k$; $\hmat{C}_k \rightarrow \mat{C}_k$
                \ENDFOR
            \ENDWHILE
            \RETURN  $\mat{A}_k, \mat{B}_k, \mat{C}_k$ and $\mat{\lambda}_k$
        \end{algorithmic}
    \caption[labelsep=period]{LL1 decomposition with non-negativity constraint}
    \label{alg:ll1_noneg}
    \end{algorithm}
}

The intuition behind the proposed common and individual feature analysis is given in the following examples.

\noindent \textbf{Example 1.}
Observe a rank-1 tensor of order 3, expressed as
\begin{equation}
    \ten{X} = \mat{a} \circ \mat{b} \circ \mat{c} = \mat{Y} \circ \mat{c} \quad \text{with } \mat{c} = \begin{bmatrix} 1 & 4 & 8 \end{bmatrix}^T
\end{equation}
According to the values of $\mat{c}$ and the definition of the outer product, the values in the first frontal slice of $\ten{X}_{(:,:,1)}$ are respectively four and eight times smaller then the values in the second $\ten{X}_{(:,:,2)}$ and third $\ten{X}_{(:,:,3)}$ frontal slices. Hence, each observation stored as the frontal slice of $\ten{X}$ exhibits the same pattern (base matrix $\mat{Y} = \mat{a} \circ \mat{b}$) that can be considered as a \textbf{common feature}. At this point, no individual information can be extracted since there is only one base matrix.

\noindent \textbf{Example 2.}
Consider a collection, $\ten{X}$, of five different color matrices stacked along the third dimension, as illustrated in \cref{fig:color_ensemble}. The tensor rank of such a 3rd order tensor (color ensemble) is three, that is, equivalent to the number of base colors (red, green, blue), which are the simplest structures from which all data can be generated through a mixing matrix $\mat{C} = [\mat{c}_R, \mat{c}_G, \mat{c}_B] \in \R^{5 \times 3}$. Thus, adopting the multi-linear notation and the RGB representation of colors, we can write
\begin{equation}
\begin{aligned}
    \ten{X} &= \ten{\Lambda} \times_1 \mat{A} \times_2 \mat{B} \times_3 \mat{C} = \ten{Y} \times_3 \mat{C}\\
            &= \ten{Y}_{(:,:,1)} \circ \mat{c}_R + \ten{Y}_{(:,:,2)} \circ \mat{c}_G + \ten{Y}_{(:,:,3)} \circ \mat{c}_B\\
    \mat{c}_R &= \begin{bmatrix} 128 & 256 & 256 & \ \ 0 \ \ & 256 \end{bmatrix}^T\\
    \mat{c}_G &= \begin{bmatrix} 128 & 256 & \ \ 0 \ \ & 256 & 128 \end{bmatrix}^T\\
    \mat{c}_B &= \begin{bmatrix} 128 & \ \ 0 \ \ & 256 & 256 & 32 \ \ \  \end{bmatrix}^T
\end{aligned}
\label{eq:cpd_colors}
\end{equation}
Here, $\ten{X} \in \R^{I \times J \times 5}$ is the original data, and $\mat{C} = [\mat{c}_R, \mat{c}_G, \mat{c}_B] \in \R^{5 \times 3}$ contains intensity values of the red, green and blue colors.
These three base colors are stored in different rank-1 frontal slices of the tensor $\ten{Y} = \ten{\Lambda} \times_1 \mat{A} \times_2 \mat{B} \in \R^{I \times J \times 3}$ and represent \textbf{common information} among the frontal slices $\ten{X}_{(:,:,n)}$. The \textbf{individual features} can be obtained by subtracting the weighted common features, to give
\begin{equation}
    \imat{X}_n = \mat{X}_n - \cmat{X} = \ten{X}_{(:,:,n)} - \sum_{k \in K_n} \alpha_k \ten{Y}_{(:,:,k)}
\end{equation}
where $K_n$ is a subset of common features for $\mat{X}_n$ with respect to the values in the n-th row of $\mat{C}$.

\includeColorEnsemble
\includeProposedALG
\subsection{LL1 decomposition with non-negativity constraint}

If a slice $\ten{Y}_{(:,:,k)}$ belongs to the  set of common features for the data sample $\ten{X}_{(:,:,n)}$, then an intuitive implication is that the corresponding value of $\mat{C}_{(n,k)}$ is positive. However, this cannot be guaranteed for a general implementation of TDs, and the non-negativity constraint should be imposed on the factor matrix $\mat{C}$, since it corresponds to the mode along which members of the ensemble are stacked together. In order to obtain more descriptive common features, we employ the LL1 decomposition from \eqref{eq:ll1}. In this way, the rank of a frontal slice $\ten{Y}_{(:,:,k)}$ is increased (see Remark 2), whereby the extraction of the common features is given by
\begin{gather}
    \ten{Y}_{(:,:,k)} = \mat{A}_k \mat{B}_k^T
\end{gather}
and requires the minimization of the cost function
\begin{gather}
    \min_{\mat{A}_k,\mat{B}_k,\mat{c}_k} \left \| \ten{X} - \sum_{k=1}^K \cpd{ \ten{\Lambda}_k ; \mat{A}_k, \mat{B}_k, \mat{c}_k } \right \|_F^2 \quad \text{s.t. } \mat{c}_k > 0
    \label{eq:ll1_cost}
\end{gather}
Notice that this problem is similar to the computation of the CPD in \eqref{eq:cpd}. Therefore, our solution is based on the ALS-CPD algorithm (we refer to \cite{kolda2009tensor} for more detail) and is summarized in \cref{alg:ll1_noneg}, where $\ast$ and $\odot$ denote respectively Khatri-Rao and Hadamard products, $(\cdot)^{\dagger}$ is the Moore–Penrose pseudoinverse, NonNegLeastSq$(\mat{X},\mat{Y})$ performs least squares on an input $\mat{X}$ and an output $\mat{Y}$, the least squares coefficients are constrained to be non-negative
\cite{NIPS2000_1861}, while repeat($\mat{X},n$) duplicates an input $\mat{X}$ \ $n$ times.

\noindent \textbf{Remark 3.} For the illustration of the proposed approach, we used a tensor of order three, however, unlike matrices the proposed approach generalises well and allows for the common and individual features to be extracted from a tensor of any order, with only one requirement that observations must be concatenated along the same mode.

    \section{Simulations and Analysis} \label{sec:simulation_results}
        \includeFaceTensorFigure

The proposed approach was employed for the classification of face images from the benchmark ORL dataset \cite{samaria1994parameterisation}. This database includes a total of the 400 grey scale images of 40 subjects in ten different illumination conditions and facial expressions. Ten sets of 40 images were created by randomly choosing one image of every subject. Six of these sets were arbitrarily selected for the training set with the remaining four forming the test set. Each group from the training set was represented as a tensor $\ten{X}^i \in \R^{112 \times 92 \times 40}$ where the images of 40 different subjects were stacked along the third dimension, as in \cref{fig:train_sample}. Their individual features were extracted by applying the proposed framework with the non-negativity constraint imposed on the mode-3 factor matrix, while the number of common features was found empirically. The classification models used were SVM, NN, QD and cKNN and were trained on the so obtained individual information. The classification scores were calculated from 100 realizations. Note that during the test stage, we used the original images in order to make a fair and realistic evaluation.

\begin{center}
\begin{tcolorbox}[tab2, fonttitle=\bfseries, fontupper=\footnotesize\sffamily, tabularx={X||Y|Y|Y|Y},title= \vspace{0.1cm} Table 1: Classification Performance in \% \vspace{0.1cm} ,boxrule=0.5pt, width=0.75\linewidth]
    \           &  SVM          & NN            & QD             & cKNN           \\[0.1cm]\hline\hline
    Original    & 83.9          & 4.35          &  \textbf{91.5} &  79.0          \\[0.2cm]
    CPD         & 91.5          & 85.8          &  89.8          &  \textbf{85.5} \\[0.2cm]
    LL1         & \textbf{94.7} & \textbf{92.2} &  86.8          &  84.3
\end{tcolorbox}
\end{center}
\cref{fig:res_orig_indiv_com} illustrates examples of images used in the experiments, and the extracted individual information and common features for the CPD and LL1 decomposition. Table 1 summarizes the performance of multi-class classification of the original images based on the corresponding individual information, extracted through the proposed method for the CPD and LL1 decompositions.

The most significant improvement can be observed for the NN based classifier. Here, the poor accuracy on the original data is associated with the high variance of the original samples and the small size of the training set which resulted in overfitting. On the other hand, the extracted individual features were of lower variance, which allowed the NN classifier to find a decision boundary that is less prone to fluctuations in the training data, leading to much higher classification accuracy.

    \section{Conclusion}
        We have proposed a novel framework for common and individual feature extraction based on the CPD and LL1 tensor decompositions with the non-negativity constraint. The multi-modal relations expressed through the outer product have been shown to play a key role in the extraction of the shared information from multi-block data. In this way, the performance of machine learning algorithms can be greatly enhanced, as the classification models use only the much lower dimensional and significantly more discriminative individual information during the training stage. Simulations have employed the ORL database of images taken from various angles, under several illumination conditions, and with different face expressions of the subjects and have achieved excellent results.
Unlike the matrix methods, the proposed method is very flexible and is not restricted to input data of a specific shared structure or images of the same dimensions.

    \bibliographystyle{IEEE}
    \bibliography{refs}

\begin{thebibliography}{10}

\bibitem{cichocki2014era}
A.~Cichocki,
\newblock ``Era of big data processing: A new approach via tensor networks and
  tensor decompositions,''
\newblock {\em arXiv preprint arXiv:1403.2048}, 2014.

\bibitem{cichocki2016tensor}
A.~Cichocki, N.~Lee, I.~Oseledets, A.~H. Phan, Q.~Zhao, and D.~P. Mandic,
\newblock ``Tensor networks for dimensionality reduction and large-scale
  optimization. {{Part}} 1: {{Low-rank}} tensor decompositions,''
\newblock {\em Foundations and Trends{\textregistered} in Machine Learning},
  vol. 9, no. 4-5, pp. 249--429, 2016.

\bibitem{cichocki2015tensor}
A.~Cichocki, D.~P. Mandic, L.~De~Lathauwer, G.~Zhou, Q.~Zhao, C.~Caiafa, and
  H.~A. Phan,
\newblock ``Tensor decompositions for signal processing applications: From
  two-way to multiway component analysis,''
\newblock {\em IEEE Signal Processing Magazine}, vol. 32, no. 2, pp. 145--163,
  2015.

\bibitem{cichocki2013tensor}
A.~Cichocki,
\newblock ``Tensor decompositions: A new concept in brain data analysis?,''
\newblock {\em arXiv preprint arXiv:1305.0395}, 2013.

\bibitem{escudero2015multiscale}
J.~Escudero, E.~Acar, A.~Fern{\'a}ndez, and R.~Bro,
\newblock ``Multiscale entropy analysis of resting-state magnetoencephalogram
  with tensor factorisations in {{Alzheimer's}} disease,''
\newblock {\em Brain Research Bulletin}, vol. 119, pp. 136--144, 2015.

\bibitem{smilde2005multi}
A.~Smilde, R.~Bro, and P.~Geladi,
\newblock ``Multi-way analysis: Applications in the chemical sciences,''
\newblock {\em Tecnometrics}, pp. 1--380, 2005.

\bibitem{kiers2001three}
H.~A. Kiers and I.~V. Mechelen,
\newblock ``Three-way component analysis: Principles and illustrative
  application,''
\newblock {\em Psychological Methods}, vol. 6, no. 1, pp. 84--110, 2001.

\bibitem{sidiropoulos2016tensor}
N.~D. Sidiropoulos, L.~De~Lathauwer, X.~Fu, K.~Huang, Evangelos~E. P., and
  C.~Faloutsos,
\newblock ``Tensor decomposition for signal processing and machine learning,''
\newblock {\em arXiv preprint arXiv:1607.01668}, 2016.

\bibitem{nguyen2015tensor}
T.~D. Nguyen, T.~Tran, D.~Q. Phung, and S.~Venkatesh,
\newblock ``Tensor-variate restricted {{Boltzmann}} machines,''
\newblock {\em In Proceedings of the 29th Conference on Artificial
  Intelligence}, pp. 2887--2893, 2015.

\bibitem{cong2015tensor}
F.~Cong, Q.~Lin, L.~Kuang, X.~Gong, P.~Astikainen, and T.~Ristaniemi,
\newblock ``Tensor decomposition of {{EEG}} signals: A brief review,''
\newblock {\em Journal of Neuroscience Methods}, vol. 248, pp. 59--69, 2015.

\bibitem{phan2013tensor}
H.~A. Phan, A.~Cichocki, P.~Tichavsk{\`y}, G.~Luta, and A.~Brockmeier,
\newblock ``Tensor completion through multiple {{Kronecker}} product
  decomposition,''
\newblock {\em In Proceedings of the 2013 IEEE International Conference on
  Acoustics, Speech and Signal Processing (ICASSP)}, pp. 3233--3237, 2013.

\bibitem{savas2007handwritten}
B.~Savas and L.~Eld{\'e}n,
\newblock ``Handwritten digit classification using higher order singular value
  decomposition,''
\newblock {\em Pattern Recognition}, vol. 40, no. 3, pp. 993--1003, 2007.

\bibitem{zink2016tensor}
R.~Zink, B.~Hunyadi, S.~Van~Huffel, and M.~De~Vos,
\newblock ``Tensor-based classification of an auditory mobile {{BCI}} without a
  subject-specific calibration phase,''
\newblock {\em Journal of Neural Engineering}, vol. 13, no. 2, pp. 1--10, 2016.

\bibitem{sorber2013optimization}
L.~Sorber, M.~Van~Barel, and L.~De~Lathauwer,
\newblock ``Optimization-based algorithms for tensor decompositions:
  {{Canonical}} polyadic decomposition, decomposition in
  {{rank-(L$_r$,L$_r$,1)}} terms, and a new generalization,''
\newblock {\em SIAM Journal on Optimization}, vol. 23, no. 2, pp. 695--720,
  2013.

\bibitem{guo2008unified}
Y.~Guo and G.~Pagnoni,
\newblock ``A unified framework for group independent component analysis for
  multi-subject {{fMRI}} data,''
\newblock {\em NeuroImage}, vol. 42, no. 3, pp. 1078--1093, 2008.

\bibitem{groves2011linked}
A.~R. Groves, C.~F. Beckmann, S.~M. Smith, and M.~W. Woolrich,
\newblock ``Linked independent component analysis for multimodal data fusion,''
\newblock {\em NeuroImage}, vol. 54, no. 3, pp. 2198--2217, 2011.

\bibitem{lock2013joint}
E.~F. Lock, K.~A. Hoadley, J.~S. Marron, and A.~B. Nobel,
\newblock ``Joint and individual variation explained {{(JIVE)}} for integrated
  analysis of multiple data types,''
\newblock {\em The Annals of Applied Statistics}, vol. 7, no. 1, pp. 523--550,
  2013.

\bibitem{zhou2016group}
G.~Zhou, A.~Cichocki, Y.~Zhang, and D.~P. Mandic,
\newblock ``Group component analysis for multiblock data: Common and individual
  feature extraction,''
\newblock {\em IEEE Transactions on Neural Networks and Learning Systems}, vol.
  27, no. 11, pp. 2426--2439, 2016.

\bibitem{zhou2015common}
G.~Zhou, A.~Cichocki, and D.~P. Mandic,
\newblock ``Common components analysis via linked blind source separation,''
\newblock {\em In Proceedings of the 2015 IEEE International Conference on
  Acoustics, Speech and Signal Processing (ICASSP)}, pp. 2150--2154, 2015.

\bibitem{kolda2009tensor}
T.~G. Kolda and B.~W. Bader,
\newblock ``Tensor decompositions and applications,''
\newblock {\em SIAM Review}, vol. 51, no. 3, pp. 455--500, 2009.

\bibitem{de2000multilinear}
L.~De~Lathauwer, B.~De~Moor, and J.~Vandewalle,
\newblock ``A multilinear singular value decomposition,''
\newblock {\em SIAM Journal on Matrix Analysis and Applications}, vol. 21, no.
  4, pp. 1253--1278, 2000.

\bibitem{de2008decompositions}
L.~De~Lathauwer,
\newblock ``Decompositions of a higher-order tensor in block terms. {{Part}}
  {{II}}: Definitions and uniqueness,''
\newblock {\em SIAM Journal on Matrix Analysis and Applications}, vol. 30, no.
  3, pp. 1033--1066, 2008.

\bibitem{NIPS2000_1861}
D.~D. Lee and H.~S. Seung,
\newblock ``Algorithms for non-negative matrix factorization,''
\newblock {\em In Proceedings of the 2000 Conference on Advances in Neural
  Information Processing Systems (NIPS)}, pp. 556--562, 2001.

\bibitem{samaria1994parameterisation}
F.~S. Samaria and A.~C. Harter,
\newblock ``Parameterisation of a stochastic model for human face
  identification,''
\newblock {\em In Proceedings of the Second IEEE Workshop on Applications of
  Computer Vision}, pp. 138--142, 1994.

\end{thebibliography}

\end{document}